\documentclass[acmsmall, screen, prologue, table]{acmart}

\def\BibTeX{{\rm B\kern-.05em{\sc i\kern-.025em b}\kern-.08emT\kern-.1667em\lower.7ex\hbox{E}\kern-.125emX}}

\setcopyright{none}

\usepackage{multirow}
\usepackage{booktabs}
\usepackage{listings}
\usepackage{amsmath}
\usepackage{amssymb}
\usepackage{arydshln}
\usepackage{mathtools}
\usepackage[noend,ruled,linesnumbered]{algorithm2e}
\usepackage{xifthen}
\usepackage{xspace}
\usepackage{textcomp}
\usepackage{caption}
\usepackage[flushleft]{threeparttable}
\usepackage{pgfplots}
\usepackage{subfig}
\usepackage{import}
\usetikzlibrary{patterns}
\pgfplotsset{width=10cm,compat=1.14}

\usepackage[toc,page]{appendix}

\newcommand{\figref}[1]{Figure~\ref{fig:#1}}

\renewcommand{\eqref}[1]{Eqn.~(\ref{Eq:#1})}

\newcommand{\tableref}[1]{Table~\ref{tbl:#1}}

\newcommand{\sectref}[1]{Section~\ref{Se:#1}}

\newcommand{\subsectref}[1]{Subsection~\ref{SubSe:#1}}

\newcommand{\subsubsectref}[1]{Subsection~\ref{SubSubSe:#1}}

\newcommand{\algref}[1]{Algorithm~\ref{Alg:#1}}

\newcommand{\lstref}[1]{Listing~\ref{Lst:#1}}
\newcommand{\lstline}[1]{(Line~\ref{lstLine:#1})}

\colorlet{sjcolor}{blue}
\colorlet{ubcolor}{teal}
\colorlet{urkcolor}{red}

\newcommand{\toolname}{{\sc LLOV}\xspace}
\newcommand{\tool}[1]{{\sc {#1}}\xspace}
\newcommand{\Omit}[1]{}
\newcommand{\EM}[1]{{\em #1}}

\newcommand{\TT}[1]{{\tt #1}}

\newcommand{\tsan}{\tool{ThreadSanitizer}}

\newcommand{\fixme}[2][]{\textcolor{red}{\textbf{URK Comment:}\ifthenelse{\isempty{#1}{}}{}{#1--}#2}}
\newcommand\redsout{\bgroup\markoverwith{\textcolor{orange}{\rule[0.5ex]{2pt}{1pt}}}\ULon}

\DeclareMathAlphabet{\mathpzc}{OT1}{pzc}{m}{it}

\definecolor{mGreen}{rgb}{0,0.6,0}
\definecolor{mGray}{rgb}{0.5,0.5,0.5}
\definecolor{mPurple}{rgb}{0.58,0,0.82}
\definecolor{backgroundColour}{rgb}{0.95,0.95,0.92}

\lstdefinestyle{CStyle}{
    commentstyle=\color{mGreen},
    keywordstyle=\color{magenta},
    numberstyle=\tiny\color{mGray},
    stringstyle=\color{mPurple},
    basicstyle=\scriptsize,
    breakatwhitespace=false,         
    breaklines=true,                 
    captionpos=b,                    
    keepspaces=true,                 
    numbers=left,                    
    numbersep=5pt,   
    showspaces=false,                
    showstringspaces=false,
    showtabs=false,                  
    tabsize=2,
    language=C++
}

\lstdefinestyle{FortStyle}{
    commentstyle=\color{mGreen},
    keywordstyle=\color{magenta},
    numberstyle=\tiny\color{mGray},
    stringstyle=\color{mPurple},
    basicstyle=\scriptsize,
    breakatwhitespace=false,         
    breaklines=true,                 
    captionpos=b,                    
    keepspaces=true,                 
    numbers=left,                    
    numbersep=5pt,   
    showspaces=false,                
    showstringspaces=false,
    showtabs=false,                  
    tabsize=2,
    language=[90]Fortran
}

\lstdefinestyle{ompStyle}{
         language=c++, 
         basicstyle=\small,
         numbers=none,            
         numberstyle=\footnotesize,
         numbersep=5pt,            
         backgroundcolor=\color{white},
         showspaces=false,             
         showstringspaces=false,       
         showtabs=false,               
         frame=single,                 
         tabsize=4,                    
         breaklines=true,              
         columns=fullflexible,
         breakautoindent=false,
         framerule=0pt,
         xleftmargin=0pt,
         xrightmargin=0pt,
         breakindent=0pt,
         resetmargins=true,
         morekeywords={pragma,omp,parallel,sections,section,single,master,critical,atomic,simd,device},
         escapeinside={(*}{*)},
    }

\makeatletter
\lstdefinelanguage{llvm}{
  morecomment = [l]{;},
  morestring=[b]", 
  sensitive = true,
  classoffset=0,
  morekeywords={
    define, declare, global, constant,
    internal, external, private,
    linkonce, linkonce_odr, weak, weak_odr, appending,
    common, extern_weak,
    thread_local, dllimport, dllexport,
    hidden, protected, default,
    except, deplibs,
    volatile, fastcc, coldcc, cc, ccc,
    x86_stdcallcc, x86_fastcallcc,
    ptx_kernel, ptx_device,
    signext, zeroext, inreg, sret, nounwind, noreturn,
    nocapture, byval, nest, readnone, readonly, noalias, uwtable,
    inlinehint, noinline, alwaysinline, optsize, ssp, sspreq,
    noredzone, noimplicitfloat, naked, alignstack,
    module, asm, align, tail, to, 
    addrspace, section, alias, sideeffect, c, gc, 
    target, datalayout, triple,
    blockaddress
    blockaddress
  },
  classoffset=1, keywordstyle=\color{purple},
  morekeywords={
    fadd, sub, fsub, mul, fmul,
    sdiv, udiv, fdiv, srem, urem, frem,
    and, or, xor,
    icmp, fcmp,
    eq, ne, ugt, uge, ult, ule, sgt, sge, slt, sle,
    oeq, ogt, oge, olt, ole, one, ord, ueq, ugt, uge,
    ult, ule, une, uno,
    nuw, nsw, exact, inbounds,
    phi, call, select, shl, lshr, ashr, va_arg,
    trunc, zext, sext,
    fptrunc, fpext, fptoui, fptosi, uitofp, sitofp,
    ptrtoint, inttoptr, bitcast,
    ret, br, indirectbr, switch, invoke, unwind, unreachable,
    malloc, alloca, free, load, store, getelementptr,
    extractelement, insertelement, shufflevector,
    extractvalue, insertvalue,
  },
  alsoletter={\%},
  keywordsprefix={\%},
}
\makeatother

\begin{document}

%
\title{\toolname: A Fast Static Data-Race Checker for OpenMP Programs}

%
\author{Utpal Bora}
\orcid{0000-0002-0076-1059}
\email{cs14mtech11017@iith.ac.in}
\author{Santanu Das}
\email{cs15mtech11018@iith.ac.in}
\author{Pankaj Kukreja}
\email{cs15btech11029@iith.ac.in}
\author{Saurabh Joshi}
\orcid{0000-0001-8070-1525}
\email{sbjoshi@iith.ac.in}
\author{Ramakrishna Upadrasta}
\orcid{0000-0002-5290-3266}
\email{ramakrishna@iith.ac.in}
\affiliation{%
  \institution{IIT Hyderabad}
  \country{India}
}
\author{Sanjay Rajopadhye}
\email{Sanjay.Rajopadhye@colostate.edu}
\affiliation{%
  \institution{Colorado State University}
  \country{USA}
}

%

\renewcommand{\shortauthors}{Bora et al.}

%
\begin{abstract}
In the era of Exascale computing, writing efficient parallel programs is indispensable and at the same time, writing sound parallel programs is very difficult. 
Specifying parallelism with frameworks such as OpenMP is relatively easy, but data races in these programs are an important source of bugs.
In this paper, we propose LLOV, a fast, lightweight, language agnostic, and static data race checker for OpenMP programs based on the LLVM compiler framework. 
We compare LLOV with other state-of-the-art data race checkers on a variety of well-established benchmarks. 
We show that the precision, accuracy, and the F1 score of LLOV is comparable to other checkers while being orders of magnitude faster. 
To the best of our knowledge, LLOV is the only tool among the state-of-the-art data race checkers that can verify a C/C++ or FORTRAN program to be data race free. 
\end{abstract}


%

%
\maketitle

\section{Introduction}\label{Se:intro}

The benefits of heterogeneous parallel programming in obtaining high performance from modern complex hardware architectures are indisputable among the scientific community. 
Although indispensable for its efficiency, parallel programming is prone to errors. 
This is crucial since programming errors could result in significant monetary losses or prove to be a risk factor where human safety systems are involved. 
Historical incidents such as Therac-25 accidents~\cite{leveson1993Therac-25}, the Ariane 5 flight 501 failure~\cite{lions1996flight}, EDS Child Support IT failure, and Knight Capital Group trading glitch~\cite{popper2012knight} have been directly attributed to software errors and testify to the need for bug detection mechanisms. 
The detection of errors a priori, therefore, could significantly reduce this risk and make programs more robust and dependable.

In this paper, we propose a solution to the problem of statically detecting data race errors in OpenMP parallel programs. 
We developed a data race detection tool based on LLVM/Clang/Flang~\cite{llvm,clang,flang} that is amenable to various languages, such as C, C++ as well as FORTRAN.
To the best of our knowledge, our work is the \textit{first static OpenMP data race detection tool based on the language independent intermediate representation of LLVM} (henceforth called LLVM-IR)~\cite{llvm-ir}.  Specifically, we make the following contributions:
\begin{itemize}
  \item Implementation of a \EM{fast}, \EM{static}, and \EM{language agnostic} OpenMP data race checker in the LLVM framework based on its intermediate representation (LLVM-IR) using the polyhedral framework Polly~\cite{grosser2012polly}.
    Our tool can also certify that a program is \EM{data race free} along with detecting race conditions in OpenMP programs.
    Moreover, our tool provides a limited support for non-affine programs using Mod/Ref information from the Alias Analyzer of LLVM.
    Additionally, the tool can be used to generate and visualize the task graph (exportable as a file in dot/gv format) of an OpenMP program.

	\item We create \textit{DataRaceBench FORTRAN}, a FORTRAN manifestation of DataRaceBench v$1.2$~\cite{liao2017dataracebench}, and release it under open source~\cite{drbForturl}. The latter is a benchmark suite consisting of programs written in C/C++ using OpenMP kernels. Our DataRaceBench FORTRAN allows for standardized evaluation of tools that can analyze FORTRAN programs for data races.

    \item We make a comparative study of well known data race checker tools on a standard set of OpenMP benchmarks. 
    We evaluate these tools on various metrics such as precision, recall, accuracy, F1 score, Diagnostic Odds Ratio and running times.
    We show that \toolname performs quite well on these metrics while completely outperforming its competitors in terms of runtime.

\end{itemize}

The rest of the paper is organized as follows:
we start with the motivation for our work in~\sectref{prelim}, and describe common data race conditions in~\sectref{Examples}. 
In~\sectref{Implement}, we discuss the verifier implementation details, our proposed algorithm, and list out the advantages of our approach over the existing dynamic tools.
\sectref{relatedWork} discusses related work in OpenMP data race detection along with their differences from our approach.
Our results and comparison with other verifiers are presented in~\sectref{results} and finally, we conclude in~\sectref{futurework}.
\section{Background and Motivation}\label{Se:prelim}
Multithreading support in hardware architectures has been very common in recent times with the number of cores per socket going up to 56 in Intel\textsuperscript{\textregistered} Xeon\textsuperscript{\textregistered} Platinum 9282 with 2 threads per core and upto 72 in Intel\textsuperscript{\textregistered} Xeon Phi\textsuperscript{\texttrademark} Processor 7290F (accelerator) with 4 threads per core. 
The Top500~\cite{top500url} November 2018 list of supercomputers comprises systems with cores per socket ranging from 6 to 260.
As these are simultaneous multithreading (SMT) systems, operating systems with support for SMT and/or Symmetric Multi Processing (SMP) can benefit from execution of a large number of threads in parallel.
The memory (not cache) is shared among all the threads in a node with either uniform or non-uniform  memory access (UMA/NUMA), enabling shared memory multithreading. 

In the past, the scientific community wrote parallel programs in C/C++ and FORTRAN using either language extensions or with APIs to run them across different nodes in a cluster or grid.
With the advent of multi-core processors, the focus shifted to a shared memory programming model, e.g., the \TT{pthreads} library/run-time system~\cite{posix2017} coupled with a vanilla language like C/C++, or a parallel language like coarray-FORTRAN or HPF.

In recent years, languages having structured parallelism such as Cilk~\cite{Blumofe1996cilk}, Julia~\cite{bezanson2017julia}, Chappel~\cite{chapel, Chamberlain:2007:PPC:1286120.1286123}, X10~\cite{charles2005x10}, and others started gaining popularity in the community.
However, the community has extensively adopted structured parallel programming frameworks, such as OpenMP~\cite{dagum1998openmp,openmpspecs}, MPI~\cite{mpi}, OpenACC~\cite{openacc}, and OpenCL~\cite{opencl} because of easy migration from legacy sequential code.
The availability of efficient runtime systems and versatile support for various architectures played a major role in popularizing these frameworks.
Amongst these, in this work we focus on the OpenMP parallel programming framework.

The OpenMP programming paradigm~\cite{dagum1998openmp,openmpspecs} introduced structured parallelism in C/C++ and FORTRAN.
It supports \emph{Single Program Multiple Data} (SMPD) programming model with multiple threads, \emph{Single Instruction Multiple Data} (SIMD) programming model within a single thread in CPUs with SIMD hardware extension, as well as SIMD among threads of a \EM{thread block} in GPUs.
OpenMP enables divide-and-conquer paradigm with tasks and nested parallelism, provides a data environment for shared memory consistency, supports mutual exclusion and atomicity, and synchronization amongst threads.

However, incorrect usage of OpenMP may introduce bugs into an application.
A common data access anomaly referred to as data race, occurs where two threads incorrectly access the same memory location, is defined formally as follows.

\begin{definition}
	[Data Race] An execution of a concurrent program is said to have a \EM{data race} when two different threads access the same memory location, these accesses are not protected by a mutual exclusion mechanism (e.g., locks), the order of the two accesses is non-deterministic 
	and one of these accesses is a write.
\end{definition}

Though compilers do ensure that OpenMP constructs conform to the syntactic and semantic specifications~\cite{openmp45spec}, \textit{none of the mainstream compilers}, such as GCC~\cite{gcc}, LLVM~\cite{llvm}, and PGI~\cite{pgi}, provide \textit{built-in data race detection} support. 
There exist dynamic tools to detect race conditions, but they either take a very long time to report races, or might miss some race conditions.
This is because these tools are dependent on the execution schedule of the threads and the program parameters.
The primary goal of our work is to provide a built-in data race detector for OpenMP parallel programs in the LLVM toolchain, using static analysis technique as discussed in~\sectref{Implement}. 

\begin{definition}
    [Team of threads] A set of OpenMP threads comprising a master thread and an optional group of sibling threads that participate in the potential execution of an OpenMP parallel region is called a \EM{team}.
    The master thread is assigned thread id zero.
\end{definition}
By default, OpenMP considers variables as \emph{shared} among all the threads in a team.

\section{Common Race conditions in OpenMP programs}\label{Se:Examples}

In this section, we will walk through, with examples, different race conditions frequently encountered in OpenMP programs.  Note that we are expansive in setting the stage here, not all the races described below can be detected by LLOV.

\subsection{Missing data sharing clauses}
\lstref{datasharingclause} shows an OpenMP worksharing construct \TT{omp parallel for} with a data race.   The program computes the sum of squares of all the elements in the matrix \TT{u}.  Here, variables \TT{temp}, \TT{i}, and \TT{j} are marked as \TT{private}, indicating that each thread will have its own copy of these variables. 
However, the variable \TT{sum}~\lstline{sum} of \lstref{datasharingclause} is not listed by any of the data sharing clauses.
Therefore, the variable \TT{sum} will be \emph{shared} among all the threads in the team.
Thus, each thread will work on the same shared variable and update it simultaneously without any synchronization, leading to a data race.

\lstref{datasharingclause_F} presents a program in FORTRAN with a data race due to a missing \TT{private} clause corresponding to the variable \TT{tmp}~\lstline{tmpwrite}.
Due to such intricacies, a programmer is prone to make mistakes and inadvertently introduce data races in the program.

Our aim is to develop techniques and a tool that understand the semantics of OpenMP pragmas and clauses with all their subtleties.

\noindent\begin{minipage}[t]{.55\textwidth}
\begin{lstlisting}[frame=single, style=CStyle, label={Lst:datasharingclause}, caption={DRB021: OpenMP Worksharing construct with data race}, language=C++, captionpos=b,basicstyle=\footnotesize\ttfamily, escapechar=|]
#pragma omp parallel for private (temp,i,j)|\label{lstLine:pragmafor}|
  for (i = 0; i < len; i++)
    for (j = 0; j < len; j++) {   
      temp = u[i][j]; 
      sum = sum + temp * temp; |\label{lstLine:sum}|
    } 
\end{lstlisting}
\end{minipage}\hfill
\begin{minipage}[t]{.4\textwidth}
\begin{lstlisting}[frame=single, style=FortStyle,label={Lst:datasharingclause_F}, caption={DRBF028: FORTRAN code with data race because of missing \TT{private} clause}, captionpos=b,basicstyle=\footnotesize\ttfamily, escapechar=|]
  !$OMP PARALLEL DO
     do i = 0, len - 1
         tmp = a(i) + i |\label{lstLine:tmpwrite}|
         a(i) = tmp |\label{lstLine:tmpread}|
     end do
  !$OMP end PARALLEL do
\end{lstlisting}
\end{minipage}

\subsection{Loop carried dependences}
OpenMP programs may suffer from race conditions due to incorrect parallelization strategies. 
Such race conditions may occur because of parallelization of loops with loop carried dependences.

For example, the loop nest in \lstref{LoopDependence} is parallel in the outer dimension~\lstline{outerloop}, but it is the inner loop~\lstline{innerloop} that is marked parallel, which has a loop carried dependence because the read of \TT{b}[i][j-1]~\lstline{bij} is dependent on write to \TT{b}[i][j]~\lstline{bij} in the previous iteration.
A correct parallelization strategy for this example would be to mark the outer loop as parallel in place of the inner loop.

As another example, \lstref{FWpar} is a  parallel implementation of Floyd-Warshall's shortest path algorithm with a \textit{benign}\footnote{A race is said to be benign (see~\cite{narayanasamy2007raceclass} for examples and analyses) if it can be formally proved that the result of the computation is unaffected by it.} race condition~\cite{basupalli2011ompverify}.  This is because all the iterations of the (parallel) inner j loop read \TT{A}[i][k] including the one where j=k, which also writes into \TT{A}[i][k].  Hence the iterations before j=k need the previous value of \TT{A}[i][k] and subsequent ones need the new, updated value.  However, if all the matrix entries are non-negative, \TT{A}[i][k] is unchanged by the assignment, therefore the race introduced by parallelizing the j loop is benign.

\noindent\begin{minipage}[t]{.42\textwidth}
\begin{lstlisting}[frame=single, style=CStyle, label=Lst:LoopDependence, caption={DRB038: Example with Loop Carried Dependence}, language=C++, captionpos=b,basicstyle=\footnotesize\ttfamily, escapechar=|]
 for (i=0;i<n;i++) { |\label{lstLine:outerloop}|
 #pragma omp parallel for |\label{lstLine:ompparfor}|
   for (j=1;j<m;j++) { |\label{lstLine:innerloop}|
     b[i][j]=b[i][j-1]; |\label{lstLine:bij}|
   }
 }
\end{lstlisting}
\end{minipage}\hfill
\begin{minipage}[t]{.53\textwidth}
\begin{lstlisting}[frame=single, style=CStyle, label=Lst:FWpar, caption={Parallel Floyd-Warshall Algorithm with a benign race condition}, language=C++, captionpos=b,basicstyle=\footnotesize\ttfamily, escapechar=|]
for (k = 1; k <= n; k++) 
  #pragma omp parallel for
  for (i = 1; i <= n; i++) 
    for (j = 1; j <= n; j++) 
      A[i][j] = 
        min(A[i][k] + A[k][j], A[i][j]);|\label{lstLine:minfunc}|
\end{lstlisting}
\end{minipage}

\subsection{SIMD races}
OpenMP supports SIMD constructs for both CPUs and GPUs. 
In CPUs, SIMD is supported with vector processing units, where the consecutive iterations of a loop can be executed in a SIMD processing unit within a single core by a single thread.
This is contrary to other loop constructs where iterations are shared among different threads.

In the example in \lstref{SIMD}, the loop is marked as SIMD parallel loop by the pragma \TT{omp simd}~\lstline{ompsimd}.
This signifies that consecutive iterations assigned to a single thread can be executed concurrently in SIMD units, called vector arithmetic logic units (ALU), within a single core rather than executing sequentially in a scalar ALU.
However, because of the forward loop carried dependence between write to \TT{a}[i+1]~\lstline{ai} in one iteration and read of \TT{a}[i]~\lstline{ai} in the previous iteration, concurrent execution of the consecutive iterations in a vector ALU will produce inconsistent results.
Dynamic data race detection tools \EM{fail to} detect race conditions in such cases as the execution happens within a single thread.\\

\noindent\begin{minipage}[t]{.45\textwidth}
\begin{lstlisting}[frame=single, style=CStyle, label=Lst:SIMD, caption={DRB024: Example with SIMD data race}, language=C++, captionpos=b,basicstyle=\footnotesize\ttfamily, escapechar=|]
#pragma omp simd |\label{lstLine:ompsimd}|
for (int i=0; i<len-1; i++){
  a[i+1] = a[i] + b[i]; |\label{lstLine:ai}|
}
\end{lstlisting}
\end{minipage}\hfill
\begin{minipage}[t]{.5\textwidth}
\begin{lstlisting}[frame=single, style=CStyle, label=Lst:NoWait, caption={DRB013: Example with data race due to improper synchronization}, language=C++, captionpos=b,basicstyle=\footnotesize\ttfamily, escapechar=|]
#pragma omp parallel shared(b, error) {|\label{lstLine:omppar}|
#pragma omp for nowait |\label{lstLine:ompnowait}|
    for(i = 0; i < len; i++) |\label{lstLine:for_nowait}|
      a[i] = b + a[i]*5; |\label{lstLine:ai_nowait}|
#pragma omp single |\label{lstLine:ompsingle}|
    error = a[9] + 1; |\label{lstLine:a9}|
  }
\end{lstlisting}
\end{minipage}

\subsection{Synchronization issues} 
Improper synchronization between threads is a common cause of data race conditions in concurrent programs. 
 OpenMP can have both explicit and implicit synchronizations associated with different constructs. 

The constructs \TT{parallel}, \TT{for}, \TT{workshare}, \TT{sections},  and \TT{single} have an implicit barrier at the end of the construct.
This ensures that all threads in the team wait for others to proceed further.
This enforcement can be overcome with the \TT{nowait} clause where threads in a team, after completion of the construct, are no longer bound to wait for the other unfinished threads.
However, improper use of the \TT{nowait} clause can result in data races as shown in  \lstref{NoWait}.
In this example, a thread executing the \TT{parallel for}~\lstline{for_nowait} will not wait for the other threads in the team because of the \TT{nowait} clause~\lstline{ompnowait}.
Threads that have finished executing the \TT{for} loop are free to continue and execute the \TT{single} construct~\lstline{ompsingle}.
Since there is a data dependence between write to \TT{a}[i]~\lstline{ai_nowait} and read of \TT{a}[9]~\lstline{a9}, it might result in a data race.

Such cases are extremely difficult to reproduce as they are dependent on the order of execution of the threads. This particular order of execution may not manifest during runtime, therefore,
making it hard for dynamic analysis tools to detect such cases.
Static analysis techniques have an advantage over the dynamic techniques in detecting race conditions for such cases.

\subsection{Control flow dependent on number of threads}
Control flow dependent on number of threads available at runtime might introduce race conditions in a parallel program.
In the example in \lstref{CtrlFlow}, race conditions will arise only when thread IDs of two or more threads in the team are multiples of 2.

\noindent
\begin{center}
    
\begin{minipage}{.60\textwidth}
\centering
\begin{lstlisting}[frame=single, style=CStyle, label=Lst:CtrlFlow, caption={Control flow dependent on number of threads}, language=C++, captionpos=b,basicstyle=\footnotesize\ttfamily, escapechar=|]
#pragma omp parallel
  if (omp_get_thread_num() % 2 == 0) {
    Flag = true;
  }
\end{lstlisting}
\end{minipage}
\end{center}

\section{Implementation Details}\label{Se:Implement}

In this section we describe the architecture, the implementation and the algorithm of our tool.

\begin{figure}[htbp]
  \centering
  \includegraphics[scale=.5]{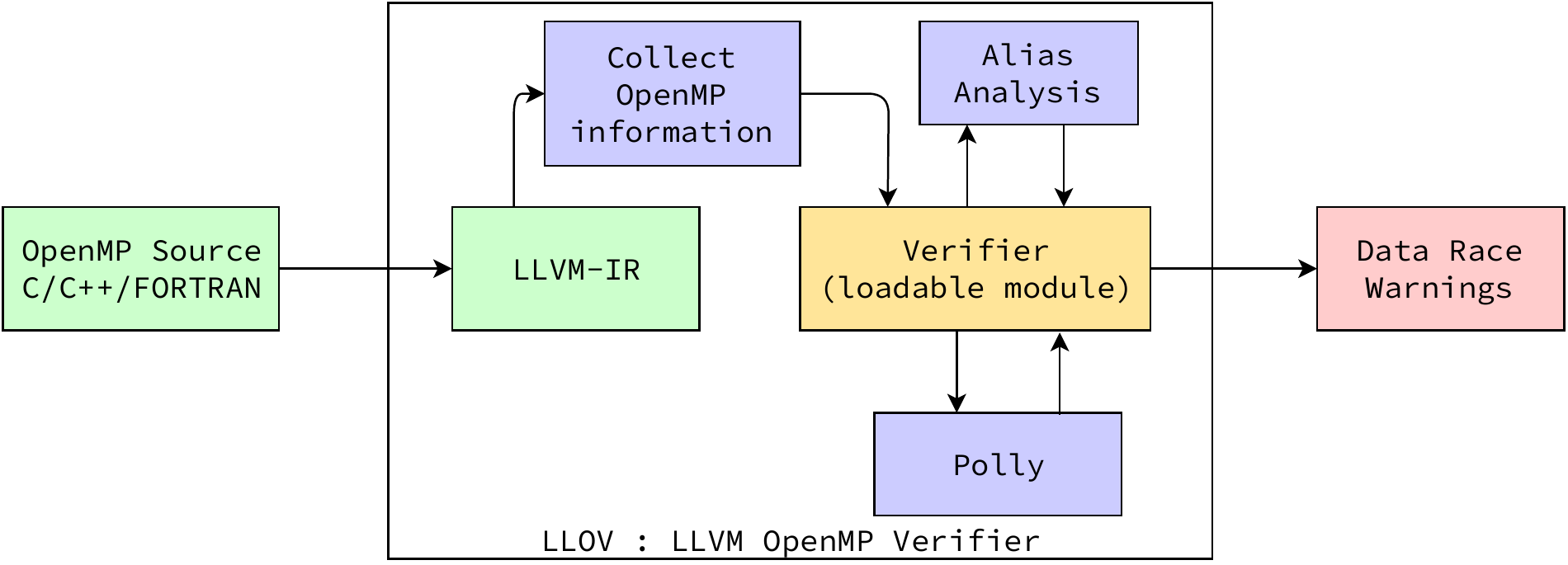}
  \caption{Flow Diagram of LLVM OpenMP Verifier (\toolname)}
   \label{fig:FlowDiag}
\end{figure}

\toolname is built on top of LLVM-IR and can analyze OpenMP programs written in C/C++ or FORTRAN. In principle, any programming language that has a stable LLVM frontend can be supported.  
LLVM-IR can be generated from C/C++ programs using the Clang~\cite{clang} frontend, and from FORTRAN programs using the Flang~\cite{flang} frontend.
The architecture of \toolname is shown in \figref{FlowDiag}.  The \toolname algorithm is primarily based on the one used in \tool{ompVerify}~\cite{basupalli2011ompverify} 

We cover the architecture and implementation details in~\subsectref{Architecture}, followed by the algorithm of our tool in~\subsectref{Algorithm}.
We discuss the advantages of our tool over dynamic race detection tools in~\subsectref{Advantages} and finally we list the limitation of the current version of our tool in~\subsectref{Limitations}.

\subsection{\toolname architecture}\label{SubSe:Architecture}
\toolname has two phases: analysis and verification.

\subsubsection{Analysis phase}
In the first phase, we analyze the LLVM-IR to collect various OpenMP pragmas and additional information required for race detection.
This analysis is necessary because OpenMP constructs are lowered to the IR by compiler frontends such as Clang~\cite{clang} and Flang~\cite{flang}.
LLVM-IR~\cite{llvm-ir} is sequential and does not have support for parallel constructs.
Hence, parallel constructs in high level languages are represented in the IR as function calls.
OpenMP pragmas are translated as function calls to the APIs of the OpenMP runtime library \EM{libomp}.

As part of this analysis, for each OpenMP construct, we collect information such as memory locations, access types, storage modifiers, synchronization details, scheduling type, etc.
An \textit{in-memory representation} of the OpenMP directive contains a directive type, a schedule (if present), variable names and types, and a list of child directives for nested OpenMP constructs. 
An illustration of our representation is shown in \lstref{OMPDirective} for the example in \lstref{NoWait} (\sectref{Examples}).
The grammar for this representation is presented in BNF form in~\lstref{PragmaRules}.

\begin{center}
\begin{minipage}[t]{0.56\textwidth}
\begin{lstlisting}[frame=single, label=Lst:OMPDirective, caption={In-memory representation of a directive}, captionpos=b,basicstyle=\scriptsize\ttfamily, escapechar=|]
Directive:  OMP_Parallel
  Variables:
    Private:   %.omp.ub = alloca i32, align 4
    Private:   %.omp.lb = alloca i32, align 4
    Shared: i32* %i
    Shared: i32* %len
    Firstprivate: i64 %vla
    Shared: i32* %a
    Shared: i32* %b
    Shared: i32* %error
    Private:   %.omp.stride = alloca i32, align 4
    Private:   %.omp.is_last = alloca i32, align 4
  Child Directives:
  1:  Directive:    OMP_Workshare_Loop
    Schedule type : Static Schedule (auto-chunked)
  2:  Directive:    OMP_Workshare_single
  3:  Directive:    OMP_Barrier
\end{lstlisting}
\end{minipage}\hfill
\begin{minipage}[t]{0.4\textwidth}
\begin{lstlisting}[frame=single, label=Lst:PragmaRules, caption={BNF of in-memory representation of directives},captionpos=b,basicstyle=\scriptsize\ttfamily]
 <Directive> ::= <Dtype> [ Sched ] 
        { <Var> } { <Directive> }
 <Dtype> ::= parallel | for | simd
        | workshare | single 
        | master | critical 
 <Var> ::= <Vtype> val
 <Vtype> ::= private | firstprivate
        | shared | lastprivate
        | reduction | threadprivate 
 <Sched> ::= [ <modifier> ] 
             [ ordered ] <Stype> 
             <chunk>
 <modifier> ::= monotonic 
        | nonmonotonic
 <Stype> ::= static | dynamic 
        | guided | auto | runtime
 <chunk> ::= positive-int-const
\end{lstlisting}
\end{minipage}
\end{center}

Reconstructing the OpenMP information from the IR has many challenges.
Not all pragmas are handled directly by the runtime.
Directives for SIMD constructs are just a hint to the optimizer that the program segment could be executed in parallel by SIMD units.
Some constructs such as worksharing \TT{for} and \TT{sections} are very similar once they are translated to LLVM-IR.
It becomes a challenge to distinguish between the two.
Recovering worksharing \TT{for} becomes challenging when the \TT{collapse} clause is used. The resulting IR has no information about the loop nest in the original program.
The data sharing clauses such as \TT{private}, \TT{shared}, \TT{firstprivate}, and \TT{lastprivate} are not explicitly annotated in the IR.
They require additional analysis and we reconstruct them using their properties.
Only \TT{reduction} and \TT{threadprivate} variables can be extracted easily.
We overcome the challenge to extract OpenMP pragmas from IR by conservatively recognizing patterns in IR that are generated from high level source code that use these pragmas.

\subsubsection{Verification phase}
\noindent \toolname checks for data races only in regions of a program marked parallel by one of the structured parallelism constructs of OpenMP listed in \tableref{pragmas}.  A crucial property of OpenMP constructs is that its specification~\cite{openmp45spec} allows only \textit{structured blocks} within a pragma.  A structured block must not contain arbitrary jumps into or out of it.  In other words, a structured block closely resembles a Single Entry Single Exit (SESE) region~\cite{dragonbook2006ed2} used in loop analyses.  To our advantage, the polyhedral framework Polly~\cite{grosser2012polly} also builds SESE regions before applying its powerful and exact dependence analysis and complex transformations.

\begin{table}
\scriptsize
\caption{Comparison of OpenMP pragma handling by OpenMP aware tools. (Y for Yes, N for No)}
\resizebox{.9\columnwidth}{!}{%

\begin{tabular}{|l|c|c|c|c|c|c|c|}
\hline
\textbf{OpenMP Pragma}         & \toolname             & \tool{ompVerify}      & \tool{PolyOMP}        & \tool{DRACO}         & \tool{SWORD}          & \tool{Archer}         & \tool{ROMP} \\ \hline
\#pragma omp parallel          & \cellcolor{green!25}Y & \cellcolor{green!25}Y & \cellcolor{green!25}Y & \cellcolor{green!25}Y & \cellcolor{green!25}Y & \cellcolor{green!25}Y & \cellcolor{green!25}Y\\
\#pragma omp for               & \cellcolor{green!25}Y & \cellcolor{green!25}Y & \cellcolor{green!25}Y & \cellcolor{green!25}Y & \cellcolor{green!25}Y & \cellcolor{green!25}Y & \cellcolor{green!25}Y\\
\#pragma omp parallel for      & \cellcolor{green!25}Y & \cellcolor{green!25}Y & \cellcolor{green!25}Y & \cellcolor{green!25}Y & \cellcolor{green!25}Y & \cellcolor{green!25}Y & \cellcolor{green!25}Y\\
\#pragma omp critical          & \cellcolor{red!25}N & \cellcolor{red!25}N   & \cellcolor{red!25}N   & \cellcolor{red!25}N   & \cellcolor{green!25}Y & \cellcolor{green!25}Y & \cellcolor{green!25}Y\\
\#pragma omp atomic            & \cellcolor{red!25}N & \cellcolor{red!25}N   & \cellcolor{red!25}N   & \cellcolor{red!25}N   & \cellcolor{green!25}Y & \cellcolor{green!25}Y & \cellcolor{green!25}Y\\
\#pragma omp master            & \cellcolor{red!25}N & \cellcolor{red!25}N   & \cellcolor{green!25}Y & \cellcolor{red!25}N   & \cellcolor{green!25}Y & \cellcolor{green!25}Y & \cellcolor{green!25}Y\\
\#pragma omp single            & \cellcolor{red!25}N & \cellcolor{red!25}N   & \cellcolor{green!25}Y & \cellcolor{red!25}N   & \cellcolor{green!25}Y & \cellcolor{green!25}Y & \cellcolor{green!25}Y\\
\#pragma omp simd              & \cellcolor{green!25}Y & \cellcolor{red!25}N   & \cellcolor{red!25}N   & \cellcolor{green!25}Y & \cellcolor{red!25}N   & \cellcolor{red!25}N   & \cellcolor{red!25}N\\
\#pragma omp parallel for simd & \cellcolor{green!25}Y & \cellcolor{red!25}N   & \cellcolor{red!25}N   & \cellcolor{green!25}Y & \cellcolor{red!25}N   & \cellcolor{red!25}N   & \cellcolor{red!25}N\\
\#pragma omp parallel sections & \cellcolor{red!25}N   & \cellcolor{red!25}N   & \cellcolor{red!25}N   & \cellcolor{red!25}N   & \cellcolor{green!25}Y & \cellcolor{green!25}Y & \cellcolor{green!25}Y\\
\#pragma omp sections          & \cellcolor{red!25}N   & \cellcolor{red!25}N   & \cellcolor{red!25}N   & \cellcolor{red!25}N   & \cellcolor{green!25}Y & \cellcolor{green!25}Y & \cellcolor{green!25}Y\\
\#pragma omp threadprivate     & \cellcolor{green!25}Y & \cellcolor{red!25}N   & \cellcolor{red!25}N   & \cellcolor{red!25}N   & \cellcolor{red!25}N   & \cellcolor{green!25}Y & \cellcolor{green!25}Y\\
\#pragma omp ordered           & \cellcolor{green!25}Y & \cellcolor{red!25}N   & \cellcolor{red!25}N   & \cellcolor{red!25}N   & \cellcolor{red!25}N   & \cellcolor{green!25}Y & \cellcolor{green!25}Y\\
\#pragma omp distribute        & \cellcolor{green!25}Y & \cellcolor{red!25}N   & \cellcolor{red!25}N   & \cellcolor{red!25}N   & \cellcolor{red!25}N   & \cellcolor{green!25}Y & \cellcolor{green!25}Y\\
\#pragma omp task              & \cellcolor{red!25}N   & \cellcolor{red!25}N   & \cellcolor{red!25}N   & \cellcolor{red!25}N   & \cellcolor{red!25}N   & \cellcolor{green!25}Y & \cellcolor{green!25}Y\\
\#pragma omp taskgroup         & \cellcolor{red!25}N   & \cellcolor{red!25}N   & \cellcolor{red!25}N   & \cellcolor{red!25}N   & \cellcolor{red!25}N   & \cellcolor{green!25}Y & \cellcolor{green!25}Y\\
\#pragma omp taskloop          & \cellcolor{red!25}N   & \cellcolor{red!25}N   & \cellcolor{red!25}N   & \cellcolor{red!25}N   & \cellcolor{red!25}N   & \cellcolor{green!25}Y & \cellcolor{green!25}Y\\
\#pragma omp taskwait          & \cellcolor{red!25}N   & \cellcolor{red!25}N   & \cellcolor{red!25}N   & \cellcolor{red!25}N   & \cellcolor{red!25}N   & \cellcolor{green!25}Y & \cellcolor{green!25}Y\\
\#pragma omp barrier           & \cellcolor{red!25}N   & \cellcolor{red!25}N   & \cellcolor{green!25}Y & \cellcolor{red!25}N   & \cellcolor{green!25}Y & \cellcolor{green!25}Y & \cellcolor{green!25}Y\\
\#pragma omp teams             & \cellcolor{red!25}N   & \cellcolor{red!25}N   & \cellcolor{red!25}N   & \cellcolor{red!25}N   & \cellcolor{red!25}N   & \cellcolor{red!25}N   & \cellcolor{red!25}N\\
\#pragma omp target            & \cellcolor{red!25}N   & \cellcolor{red!25}N   & \cellcolor{red!25}N   & \cellcolor{red!25}N   & \cellcolor{red!25}N   & \cellcolor{red!25}N   & \cellcolor{red!25}N\\
\#pragma omp target map        & \cellcolor{red!25}N   & \cellcolor{red!25}N   & \cellcolor{red!25}N   & \cellcolor{red!25}N   & \cellcolor{red!25}N   & \cellcolor{red!25}N   & \cellcolor{red!25}N\\
\hline

\end{tabular}
}
\label{tbl:pragmas}
\end{table}

The polyhedral framework is based on \textit{exact dependence (affine) analysis}~\cite{feautrier91dataflow}, by which the dependence information can be expressed as (piecewise, pseudo-) affine functions.
Polly~\cite{grosser2012polly, polly} is the polyhedral compilation engine in LLVM framework.
Polly relies on the Integer Set Library (ISL)~\cite{verdoolaege2010isl} to perform exact dependence analysis, using which Polly performs transformations such as loop tiling, loop fusion, and outer loop vectorization.
Dependences are modelled as ISL relations and transformations are performed on integer sets using ISL operations.
The input to Polly is serial code that is marked as Static Control Part (SCoP) and the output is tiled or parallel code enabling vectorization.

\toolname is built using the Polly infrastructure, but does not use its transformation capabilities.
Our primary goal is to perform analyses, whereas Polly is designed for complex parallelizing or locality transformations followed by polyhedral code generation.
The input to \toolname is \textit{explicitly parallel code} that uses the structured parallelism of OpenMP. 
With an assumption that the input code has a serial schedule\footnote{This assumption is trivial to prove because of the input C language semantics.}, \toolname calculates its dependence information by using the dependence analyzer of Polly.
Using this dependence information, \toolname then analyzes the parallel constructs and checks the presence or absence of data races.
Consequently, \toolname can deterministically state whether a program has data races, or whether it is race free for an affine subset of programs.

Polly was designed~\cite{grosser2011polly, grosser2012polly} as an automatic parallelization pass using polyhedral dependence analysis.
Its analysis and transformation phases were closely coupled and the analyses were not directly usable from outside Polly, neither by other analyses nor by optimization passes in LLVM.
In particular, SCoP detection and dependence analysis of Polly was not directly accessible from analyses in LLVM.

We modified and extended the analysis phase of Polly in such a way that its internal data structures become accessible from LLVM\footnote{The initial version of the implementation for exposing Polly's dependence analysis information to LLVM was published in Polly (as part of \textit{Google Summer of Code} 2016 project) \textit{``Polly as an analysis pass in LLVM''}~\cite{utpal2016gsoc}.}.
Other changes involved modifying the dependence analysis of Polly so that its Reduced Dependence Graph (RDG) could be computed \textit{on-the-fly} for a function, thereby reducing the analysis time for \toolname. Polly can detect and model only sequential programs. Hence, it does not support OpenMP programs.
We incorporated changes to the SCoP detection to model OpenMP programs assuming a sequential schedule.
The OpenMP parallel LLVM-IR contains runtime library calls to change the loop bounds at runtime, which makes the program non-affine.
We mitigate this problem by resetting the loop bounds to the original values.

The two phases of \toolname are not tightly coupled, meaning the verification phase is separate from the analysis phase.
The advantage of having a two phase design is that the verifier could easily be plugged with another analysis phase for other parallel programming APIs such as Intel TBB~\cite{inteltbburl, inteltbb}, OpenCL~\cite{opencl} once they have a translator to LLVM-IR.

\subsection{Race detection algorithm}\label{SubSe:Algorithm}

First we cover the race detection for affine regions which relies on Polly, followed by the race detection for non-affine regions which relies on the Alias Analysis of LLVM.

\begin{minipage}[t]{.46\textwidth}
\begin{algorithm}[H]
\footnotesize
\KwIn{Loop $L$}
\KwOut{True/False}
\SetKwFunction{FMain}{isRaceFree}
  \SetKwProg{Fn}{Function}{:}{}
  \Fn{\FMain{$L$}}{
  $SCoP$ = ConstructSCoP($L$) \;
  $RDG$ = ComputeDependences($SCoP$) \;
  $depth$ = GetLoopDepth($L$) \;
  \eIf{isParallel(RDG, depth)}{
   \textit{// Program is race free.} \\
   return True \;
   }{
   \textit{// Data Race detected.} \\
   return False \;
  }
  \textbf{return} result
 }
 \textbf{End Function}
 \caption{Race Detection Algorithm}
 \label{Alg:race1}
\end{algorithm}
\end{minipage}\hfill
\begin{minipage}[t]{.5\textwidth}
\begin{algorithm}[H]
\footnotesize
\SetAlgoLined
\KwIn{$RDG$, Loop-depth $dim$}
\KwOut{True/False}
\SetKwFunction{FMain}{isParallel}
  \SetKwProg{Fn}{Function}{:}{}
  \Fn{\FMain{$RDG$, $dim$}}{
  \eIf{RDG is Empty}{
    \textbf{return} True \;
  }{
  Flag = True\;
  \While{Dependence D in RDG}{
  $D'$ = Project Out all dimensions except $dim$ from $D$ \;
  \eIf{D' is Empty}{
   \textbf{continue} \;
   }{
   Flag = False \;
   \textbf{break} \;
  }
  }
  \textbf{return} Flag \;
 }
 }
 \textbf{End Function}
\textbf{\caption{Algorithm to check parallelism}
 \label{Alg:race2}
} 
 \end{algorithm}
\end{minipage}

\subsubsection{Race detection in affine regions}\label{SubSubSe:RaceAffine}

In the analysis phase, we gather information provided by OpenMP's structured parallelism constructs. 
We model a section of code, marked as parallel by one of the OpenMP constructs listed in~\tableref{pragmas}, as static affine control parts (SCoPs) in the polyhedral framework.
Our race detection algorithm runs only on sections of a program marked as parallel by one of these pragmas.
This reduces analysis time of \toolname as it can avoid performing dependence analysis on the sequential fragments of the program.

For each SCoP, we query the race detection \algref{race1} to check for the presence of dependences. 
A race condition is flagged when the set of the memory accesses in the reduced dependence graph (RDG) and the set of the shared memory accesses within the SCoP is not disjoint.
When dependences are absent, the SCoP is parallel and the corresponding program segment is guaranteed to be data race free. 
Hence, we can verify---fully statically---the \textit{absence of data race} in a program. 

\begin{center}
\begin{minipage}[t]{.5\textwidth}
\begin{lstlisting}[frame=single, style=CStyle, label=Lst:Math, caption={Two dimensional loop nest}, language=C++, captionpos=b,basicstyle=\footnotesize\ttfamily, escapechar=|, belowcaptionskip=-1pt]
for (i=0;i<m;i++) { |\label{lstLine:outerloop-par}|
  for (j=1;j<n;j++) { |\label{lstLine:innerloop-ser}|
    b[i][j]=b[i][j-1]; |\label{lstLine:MA-bij}|
  }
}
\end{lstlisting}
\end{minipage}
\end{center}

\begin{figure}[H]
\subfloat[Dependence Polyhedra]{%
\begin{tikzpicture}[scale=0.4]
\centering
\draw[step=1cm,gray,very thin] (-.9,-.9) grid (6.9,6.9);
\draw[thick,->] (0,0) -- (6.5,0) node[anchor=north west] {i};
\draw[thick,->] (0,0) -- (0,6.5) node[anchor=south east] {j};
\foreach \x in {0,1,2,3,4,5,6}
   \draw (\x cm,1pt) -- (\x cm,-1pt) node[anchor=north] {$\x$};
\foreach \y in {0,1,2,3,4,5,6}
    \draw (1pt,\y cm) -- (-1pt,\y cm) node[anchor=east] {$\y$};
\foreach \i in {1,2,3,4,5,6} {
    \foreach \j in {1,2,3,4,5,6} {
        \node at (\i,\j) {};
        \node at (\j,\i) {};
        \draw[thin,red,->] (\i,\j) -- (\i,\j-1);
    }
}
\label{fig:rdg}
\end{tikzpicture}
}
\hspace*{5mm}
\subfloat[Projection of \ref{fig:rdg} on i-dimension]{%
\begin{tikzpicture}[scale=0.4]
\centering
\draw[step=1cm,gray,very thin] (-.9,-.9) grid (6.9,6.9);
\draw[thick,->] (0,0) -- (6.5,0) node[anchor=north west] {i};
\draw[thick,->] (0,0) -- (0,6.5) node[anchor=south east] {j};
\foreach \x in {0,1,2,3,4,5,6}
   \draw (\x cm,1pt) -- (\x cm,-1pt) node[anchor=north] {$\x$};
\foreach \y in {0,1,2,3,4,5,6}
    \draw (1pt,\y cm) -- (-1pt,\y cm) node[anchor=east] {$\y$};
\foreach \i in {0,1,2,3,4,5,6} {
    \foreach \j in {0} {
        \fill[very thick, red] (\i,\j) circle (2pt);
    }
}
\label{fig:projecti}
\end{tikzpicture}
}
\hspace*{5mm}
\subfloat[Projection of \ref{fig:rdg} on j-dimension]{%
\begin{tikzpicture}[scale=0.4]
\centering
\draw[step=1cm,gray,very thin] (-.9,-.9) grid (6.9,6.9);
\draw[thick,->] (0,0) -- (6.5,0) node[anchor=north west] {i};
\draw[thick,->] (0,0) -- (0,6.5) node[anchor=south east] {j};
\foreach \x in {0,1,2,3,4,5,6}
   \draw (\x cm,1pt) -- (\x cm,-1pt) node[anchor=north] {$\x$};
\foreach \y in {0,1,2,3,4,5,6}
    \draw (1pt,\y cm) -- (-1pt,\y cm) node[anchor=east] {$\y$};
\foreach \i in {0} {
    \foreach \j in {1,2,3,4,5,6} {
        \fill[red] (\i,\j) circle (1pt);
        \draw[very thick,red,->] (\i,\j) -- (\i,\j-1);
    }
}
\label{fig:projectj}
\end{tikzpicture}
}
\caption{Dependence polyhedra and its projections on i \& j dimensions}
\label{fig:method}
\end{figure}
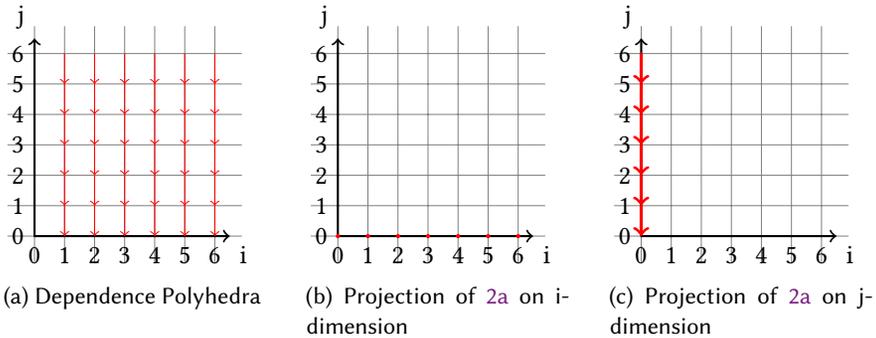

The polyhedral representation of the affine static control program in~\lstref{Math} consists of an iteration domain (\textbf{I}), an execution order called schedule (\textbf{S}), and an access function (\textbf{A}) mapping iteration number to memory accesses.
The RDG (\textbf{D}) is computed using this information. The RDG is shown graphically in ~\figref{rdg}.

\begin{center}
\begin{minipage}[t]{\textwidth}
\begin{tabular}{lc}
Iteration Domain : &
$\textbf{I} = \{ \texttt{S0}(i, j) :  0 \leq i \leq m-1  \land 1 \leq j \leq n-1 \} $\\
Schedule : &
$ \textbf{S} =\{ \texttt{S0}(i, j) \rightarrow (i, j) \} \cap_{dom} \textbf{I} $\\
Access Map : &
$ \textbf{A}=\{ \texttt{S0}(i, j) \rightarrow \texttt{M}(i, j) ;  \texttt{S0}(i, j) \rightarrow \texttt{M}(i, j-1)\} $\\
Dependences : &
$ \textbf{D} =\{ \texttt{S0}(i,j) \rightarrow (i,j-1) : 0 \leq i \leq m-1 \land 1 \leq j \leq n-1 \} $\\
\end{tabular}
\end{minipage}
\end{center}

\figref{projecti} shows the projection of the RDG on i-dimension, which results in vectors with zero magnitude as represented by red dots. This signifies that the loop is parallel in the i-dimension.
\figref{projectj} shows the projection of the RDG on j-dimension, which are vectors of unit magnitude.
Non-zero magnitude means that this dimension is not parallel.

Time complexity of the race detection algorithm is exponential in the number of inequalities present since our approach is based on Fourier\textendash Motzkin elimination.

\vspace{-2em}

\subsubsection{Race detection in non-affine regions}\label{SubSubSe:RaceNonAffine}
In addition to race detection in affine regions that can be exactly modelled by Polly, we use LLVM's Alias Analysis (AA) to conservatively analyze non-affine regions that cannot be modelled by Polly.

We use the Mod/Ref information from the Alias Analysis engine of LLVM~\cite{llvm-aa} to analyze whether a shared memory location is read (Ref) or modified (Mod) by an instruction.
The AA engine provides generic helper functions to return the Mod/Ref information for a memory location and an instruction of one of the following types: callsite, load, store, atomic read-write, invoke, etc.
The Mod and Ref bits are set for an instruction if the execution of the instruction might modify or reference the specified memory location.

If an operation inside a parallel region on the specified memory location is not protected by locks and Mod/Ref is set, \toolname flags a race signaling a potential data-race condition.
The AA race checks are invoked only when the region cannot be modelled by Polly as an affine region.
The Mod/Ref analysis of LLVM is conservative, and can lead to \toolname producing false positive races.
Thus \toolname provides a limited support of non-affine programs.

These AA-based checks for race detection in \toolname are enabled by default; we also provide a flag (\EM{-openmp-verify-disable-aa}) which can be used to disable these and run only the polyhedral verifier of \toolname.

\subsection{Advantages over dynamic race detection tools}\label{SubSe:Advantages}
Our static race detection tool \toolname has several advantages over state-of-the-art dynamic race detection tools. 

\subsubsection{Detects races in SIMD constructs}
\noindent \toolname can detect SIMD races within a single thread, such as those shown in \lstref{SIMD}.  
It can detect parallelism of the SIMD loop within the loop nest, but even when the loop is not parallel, there is a possibility of data races due to concurrent execution of consecutive iterations in the SIMD units within a single core.
Dynamic race detection tools are based on different techniques such as vector clocks, happens-before relations, locksets, monitors, offset-span-labels, etc., and fail to detect such race conditions present within a single thread.

\subsubsection{Independent of runtime thread schedule}
Being a static analysis tool, \toolname has the added advantage that the race detection is not dependent on the order of the execution of the threads.  This is a major drawback of the dynamic tools because they need to be run multiple times for each specific number of threads to detect races dependent on the execution order.

\subsubsection{Independent of input size}
\noindent \toolname can handle parametric array sizes and loop bounds.
Since we solve the problem with parametric integer programming~\cite{feautrier1988parametric}, there is no limitation on the input size, provided the control flow is not affected by it.
On the other hand, the dynamic data race detection tools need to be run multiple times for each program parameter to capture races.
It is computationally not feasible to cover all the possible input sizes and hence, a dynamic tool can never be complete.

\subsubsection{Independent of number of threads}
Our analysis is not dependent on the number of threads available during runtime.
However, all known dynamic tools have to be run multiple times with different numbers of threads to detect races.
Hence, dynamic tools might miss out race conditions when the number of runtime threads is small.

\subsection{Limitations of \toolname}\label{SubSe:Limitations}
\toolname is in active development and in the current version, we attempted to cover the frequently used OpenMP v4.5 pragmas.
In the current version, \toolname does not provide support for the OpenMP constructs for synchronization, device offloading, and tasking.
Function calls within an OpenMP construct are handled by \toolname only if the function is inlined.
Also, since our tool is primarily based on the polyhedral framework, its application is limited by the affine restrictions.
However, \toolname provides a limited support for non-affine programs using Mod/Ref analysis as discussed in~\subsubsectref{RaceNonAffine}.
Programs with dynamic control flow and irregular accesses (like $a[b[i]]$) fall outside the purview of the polyhedral framework.
We are working on extending the analysis of our tool on such non-affine programs, but that is beyond the current scope.

\toolname may produce False Negatives when there is a race due to a dependence across two static control parts (SCoPs) of a program. One such category of programs is the presence of \TT{nowait} clause in a worksharing \TT{for} construct.

\toolname can produce False Positives for programs with explicit synchronizations with barriers and locks.
Programs generated by automatic parallelization tools such as PolyOpt~\cite{polyopturl} and PLuTo~\cite{plutourl} will have complex loop bounds and are difficut to model precisely.
\toolname might produce FP for some of these automatically generated tiled or parallel kernels.

Handling \TT{sections} construct in LLVM-IR is a challenge, as the resulting IR is similar to worksharing \TT{for} construct but with a control dependence on the number of threads.
Hence \toolname might produce FP/FN cases instead of showing a diagnostic message stating that the construct is outside its purview.

\section{Related Work}\label{Se:relatedWork}

There has been extensive work on data race detection in parallel programs; 
many static, dynamic, and hybrid analyses approaches have been proposed.
Mellor-Crummey et. al.~\cite{mellorcrummey1991offset-span} proposed race detection in fork-join multithreaded programs using Offest-Span labelling of the nodes.
\tool{Eraser}~\cite{savage1997eraser} proposed lockset based approach for race detection.
Most of the earlier works have focused on \TT{pthread} based programs, while recent works such as \tool{ompVerify}~\cite{basupalli2011ompverify}, \tool{Archer}~\cite{atzeni2016archer}, \tool{ROMP}~\cite{gu2018romp}, \tool{SWORD}~\cite{atzeni2018sword}, \tool{DRACO}~\cite{ye2018draco}, and \tool{PolyOMP}~\cite{chatarasi2016extended} have targeted OpenMP programs.

In the following subsections, we discuss the state-of-the-art tools and categorize them based on their analyses and their approaches.

\subsection{Static Tools}
There are multiple static data race detectors in the literature.
Common techniques used for static analyses are lockset based approach or modeling race condition as a linear programming problem (integer-linear or parametric integer-linear), and appropriately using an ILP or SMT solver to check for its satisfiability.
Here, we briefly cover the state-of-the-art static data race detection tools for OpenMP programs.
We limit the discussion to OpenMP race detection tools only.

\tool{ompVerify}~\cite{basupalli2011ompverify} is a polyhedral model based static data race detection tool that detects incorrectly specified \TT{omp parallel for} constructs. 
\tool{ompVerify} computes the Polyhedral Reduced Dependency Graph (PRDG) to reason about possible violations of true dependences, write-write conflicts, and stale read problems in the OpenMP \TT{parallel for} construct. 
Although our approach is inspired by \tool{ompVerify}, we have implemented a different algorithm to detect parallelism of loop nests.
And, while \tool{ompVerify} can handle only the \TT{omp parallel for} construct, the coverage of \toolname is \textit{much wider}; it can handle many more pragmas as listed in \tableref{comparison}.
Moreover, the prototype implementation of  \tool{ompVerify} is in Eclipse CDT/CODAN framework using the AlphaZ~\cite{yuki2012alphaz} polyhedral framework whereas \toolname is based on the widely used LLVM compiler infrastructure.
Finally, \tool{ompVerify} works at the AST level, whereas \toolname works on the language independent LLVM-IR level making it applicable to multiple languages.
LLOV also has a limited support for non-affine regions which \tool{ompVerify} does not have.

\tool{DRACO}~\cite{ye2018draco} is a static data race detection tool based on the Polyhedral model and is built on the ROSE compiler framework~\cite{quinlan2011rose, SchordanQuinlan2003rose}.
One significant advantage of \toolname over \tool{DRACO} is that \toolname is based on LLVM-IR and is language agnostic, while \tool{DRACO} is limited only to the C family of languages that can be compiled by the ROSE compiler.

\tool{PolyOMP}~\cite{chatarasi2016static,chatarasi2016extended} is a static data race detection tool based on the polyhedral model.
\tool{PolyOMP}~\cite{chatarasi2016static} uses an extended polyhedral model to encode OpenMP loop nest information as constraints and uses the Z3~\cite{moura2008z3} solver to detect race conditions. 
The extended version of \tool{PolyOMP}~\cite{chatarasi2016extended} uses May-Happen-in-Parallel analysis in place of Z3 to detect race conditions.
In contrast, \toolname uses RDG (Reduced Dependence Graph) to determine parallelism in a region and infer race conditions based on the presence of data dependences.

Other static analysis tools like 
\tool{Relay}~\cite{voung2007relay}, 
\tool{Locksmith}~\cite{pratikakis2011locksmith} and
\tool{RacerX}~\cite{engler2003racerx}
use \tool{Eraser}'s~\cite{savage1997eraser} lockset algorithm and can detect races in \TT{pthread} based C/C++ programs.
\tool{RacerD}~\cite{blackshear2018racerd} is a static analysis tool for Java programs.

\subsection{Dynamic Tools}
Various dynamic race detection techniques have been proposed in the literature.
The well known among them are based on Locksets~\cite{savage1997eraser}, Happens-before~\cite{lamport1978happens-before} relations,
and Offset-Span labels~\cite{mellorcrummey1991offset-span}.

\tool{Archer}~\cite{atzeni2016archer} uses both static and dynamic analyses for race detection. 
It uses happens-before relations~\cite{lamport1978happens-before, serebryany2009threadsanitizer} which enforces multiple runs of the program to find races. 
\tool{Archer} reduces the analysis space of \TT{pthread} based tool \tool{TSan-LLVM} by instrumenting only parallel sections of an OpenMP program.
As \tool{Archer} uses shadow memory to keep track of each memory access, memory requirement still remains the problem for memory bound programs.

In the analysis phase, \tool{Archer} uses Polly to get the dependent loads and stores for a function and blacklist a section of code in the absence of dependences.
However, presence of dependence in a loop nest need not result in a data race.
\begin{center}
\begin{minipage}[t]{.80\textwidth}
\begin{lstlisting}[frame=single, style=CStyle, label=Lst:ArcherDiff, caption={Loop nest with Loop Carried Dependence but without any data race}, language=C++, captionpos=b,basicstyle=\footnotesize\ttfamily, escapechar=|]
 #pragma omp parallel for |\label{lstLine:ompparforA}|
 for (i=0;i<n;i++) { |\label{lstLine:outerloopA}|
   for (j=1;j<m;j++) { |\label{lstLine:innerloopA}|
     b[i][j]=b[i][j-1]; |\label{lstLine:bijA}|
   }
 }
\end{lstlisting}
\end{minipage}
\end{center}
\vspace{-.5em}
The example in \lstref{ArcherDiff} consists of a loop nest where the outer loop is parallel but the inner loop has a loop carried dependence.
However, this is a valid parallelization strategy since only the outer loop is marked parallel.
With only loads and stores information, \tool{Archer} will not be able to statically blacklist such code.
It will have to rely on its dynamic analysis using \tool{TSan-LLVM}.
Moreover, the version of \tool{Archer} (git master branch commit hash fc17353) used in our experiments completely disabled static analysis and does not use Polly at all.
\tool{Archer} only uses OMPT~\cite{eichenberger2013ompt} callbacks to instrument the code with happens-before annotations for \tool{TSan-LLVM}.

Since LLOV checks for parallelism at each level of the loop nest as stated in~\subsectref{Algorithm}, hence it can statically detect the loop nest as data race free.

\tool{SWORD}~\cite{atzeni2018sword} is a dynamic tool based on operational semantic rules and uses OpenMP tools framework OMPT~\cite{eichenberger2013ompt}.
\tool{SWORD} uses locksets to implement the semantic rules by taking advantage of the events tracked by OMPT.
\tool{SWORD} logs runtime traces for each thread consisting of all the OpenMP events and memory accesses using OMPT APIs.
In the second offline phase, it analyzes the traces for concurrent threads using Offset-Span labels and detects unsynchronized memory accesses in two concurrent threads. 
If no synchronization is used on a common memory access, data race condition is flagged. 
\tool{SWORD} cannot detect races in OpenMP SIMD, tasks and target offloading constructs. 

\tool{ROMP}~\cite{gu2018romp} is a dynamic data race detection tool for OpenMP programs. 
\tool{ROMP} maintains access history for each memory access.
An access history consists of the access event, access type, and any set of locks associated with the access. 
\tool{ROMP} constructs a task graph for the implicit and explicit OpenMP tasks and analyzes concurrent events.
If an access event is concurrent and the memory is not protected by mutual exclusion mechanisms, \tool{ROMP} flags data race warnings.
\tool{ROMP} builds upon the offset-span-labels of OpenMP threads and constructs task graphs to detect races.

\tool{Helgrind}~\cite{valgrind2007helgrind} is a dynamic data race detection tool built in Valgrind framework~\cite{valgrind2003url} for C/C++ multithreaded programs.
\tool{Helgrind} maintains \textit{happens-before} relations for each pair of memory accesses and forms a directed acyclic graph (DAG). 
If there is no path from one location to another in the happens-before graph, data race is flagged.

\tool{Valgrind DRD}~\cite{valgrind2007drd} is another dynamic race detection tool in Valgrind. 
It can detect races in multithreaded C/C++ programs.
It is based on happens-before relations similar to \tool{Helgrind}.

\tool{ThreadSanitizer}~\cite{serebryany2009threadsanitizer} is a dynamic data race detection tool based on Valgrind for multi-threaded C, C++ programs. 
It employs a hybrid approach of happens-before annotations and maintains locksets for read and write operations on shared memory. 
It maintains a state machine as metadata called shadow memory. 
It reports a race condition when two threads access the same memory location and their corresponding locksets are disjoint.
Because of the shadow memory requirement for each memory access, \tool{ThreadSanitizer}'s memory requirement grows linearly with the amount of memory shared among threads. 
Binary instrumentation also increases the runtimes by around 5x-30x~\cite{serebryany2011tsan-llvm}.

\tool{TSan-LLVM}~\cite{serebryany2011tsan-llvm} is based on \tsan~\cite{serebryany2009threadsanitizer}.
\tool{TSan-LLVM} uses LLVM to instrument the binaries in place of Valgrind.
\tool{TSan-LLVM} instrumented binaries incur less runtime overhead compared to \tool{ThreadSanitizer}. 
However, it still has similar memory requirements and remains a bottleneck for larger programs. 

\tool{Intel Inspector}~\cite{intelinspector} is a commercial, dynamic data race detection tool for C, C++ and FORTRAN programs.

There are other dynamic analysis tools like \tool{Eraser}~\cite{savage1997eraser}, \tool{FastTrack}~\cite{flanagan2009fasttrack}, \tool{CoRD}~\cite{kasikci2012cord}, and \tool{RaceTrack}~\cite{yo2005racetrack} that use the lockset algorithm to detect races in parallel programs.
\tool{IFRit}~\cite{effinger-Dean2012ifrit} is a dynamic race detection tool based on Interference Free Regions(IFR).
Since they are not specific to OpenMP, we have not discussed them here. \\

\noindent \EM{OpenMP aware tools:}
Majority of the tools are either POSIX thread based or are specific to race detection in inherent parallelism of various programming languages, such as Java, C\#, X10, and Chappel.
The tools \tool{ompVerify}~\cite{basupalli2011ompverify}, \tool{Archer}~\cite{atzeni2016archer}, \tool{PolyOMP}~\cite{chatarasi2016extended}, \tool{DRACO}~\cite{ye2018draco}, \tool{SWORD}~\cite{ye2018draco}, and \tool{ROMP}~\cite{gu2018romp} are the only ones that exploit the intricate details of structured parallelism in OpenMP.

\tool{OMPRacer}~\cite{swain2020ompracer} is another recent LLVM based tool that was announced during the course of publication of our work.
\tool{OMPRacer} relies on alias analysis, happens-before relations and locksets to statically detect races in OpenMP programs. 
The published version of \tool{OMPRacer}~\cite{swain2020ompracer} has a comparison with an initial version of \toolname. 
The comparison is limited to only C/C++ benchmarks, with no mention of comparison using FORTRAN benchmarks.
Comparing with their published numbers, the current version of \toolname outperforms \tool{OMPRacer} in precision, recall, and accuracy on DataRaceBench v1.2 benchmark.\\

\noindent \EM{Polyhedral model based static analysis tools:}
To the best of our knowledge, \tool{ompVerify}~\cite{basupalli2011ompverify}, \tool{DRACO}~\cite{ye2018draco}, \tool{PolyOMP}~\cite{chatarasi2016extended}, and \toolname are the only tools for race detection in OpenMP programs that are based on the polyhedral framework.
To put it theoretically, these are the only tools that use the exact dependence analysis of polyhedral compilation, which is crucial for  the exact analysis of a large class of useful loop programs~\cite{benabderrahmane2010widely}.

\toolname is different from these tools as it works on the LLVM-IR and collects OpenMP pragmas that have been lowered to library calls. This makes \toolname language independent.
Also, the analysis phase of \toolname could be used for other purposes, like generating task graphs from the LLVM-IR.\\

\noindent \EM{Race detection in OpenMP programs for GPUs:}
Multitude of works
\cite{cuda-memcheck,zheng2011grace,zheng2014gmrace,li2010smt,li2012gklee,li2012symrace,li2017ld,peng2018curd,li2014symrace,betts2012gpuverify}
have investigated the problem of verification of Compute Unified Device Architecture (CUDA) programs.
However, not much work has gone into verification of OpenMP device offloading constructs.
This might be due to lack of complete support for OpenMP device offloading to CUDA devices by the mainstream compilers.
The recent work from Barua et al.~\cite{barua2019ompsan} performs verification of the OpenMP host-device data mapping with the \EM{map} clause.
Arm DDT~\cite{arm-ddt} is a commercial debugging tool that supports debugging of OpenMP and CUDA programs.\\

\noindent Recent works by Liao et. al.~\cite{liao2017dataracebench,liao2017dataracebenchv12} and Lin et. al.~\cite{lin2019regressionrace} compare different race detection tools using the  DataRaceBench~\cite{liao2017dataracebench} benchmark.
In~\sectref{results}, we show comparison of our tool \toolname with the other race detection tools on DataRaceBench, as well as two other benchmarks in greater detail. 
\section{Experimental Results}\label{Se:results}

In this section we describe our experimental setup, provide details on our experiments, and compare the results with other tools on a set of benchmarks.

\subsection{Experimental Setup}

\begin{table}[t]
\centering
\small
\caption{Race detection tools with the version numbers used for comparison}
\label{tbl:Tools}
\begin{tabular}{ lcc }
\toprule
\textbf{Tools} & \textbf{Source} & 
\textbf{Version / Commit} \\
\midrule
\tool{Helgrind}~\cite{valgrind2007helgrind} & Valgrind & 3.13.0 \\
\tool{Valgrind DRD}~\cite{valgrind2007drd} & Valgrind & 3.13.0 \\
\tool{TSan-LLVM}~\cite{serebryany2009threadsanitizer} & LLVM & 6.0.1 \\
\tool{Archer}~\cite{atzeni2016archer} & git master branch & fc17353 \\
\tool{SWORD}~\cite{atzeni2018sword} & git master branch & 7a08f3c \\
\tool{ROMP}~\cite{gu2018romp} & git master branch & 6a0ad6d \\
\bottomrule
\end{tabular}
\end{table}

We compare \toolname with the state-of-the-art data race detection tools as listed in \tableref{Tools}.
Some of the relevant tools were left out of our experimentation either because of their unavailability or due to their inability to handle OpenMP programs.
We had issues setting up \tool{ROMP}~\cite{gu2018romp} and hence could not consider it for our comparison.
\tool{ompVerify}~\cite{basupalli2011ompverify} is a prototype implementation in the Eclipse CDT/CODAN framework and can detect races in \EM{omp parallel for} constructs only.
Neither the binary nor source code for  \tool{PolyOmp}~\cite{chatarasi2016extended} and  \tool{DRACO}~\cite{ye2018draco} are available in the open.
We did not consider \tool{Intel Inspector} due to its proprietary nature and licensing issues\footnote{We could not procure an educational license for Intel ICC.}.

For evaluation we chose DataRaceBench v1.2~\cite{liao2017dataracebenchv12, drbv12url}, DataRaceBench FORTRAN~\cite{drbForturl} (a FORTRAN implementation of DataRaceBench v1.2), and OmpSCR~v2.0~\cite{ompSCR} benchmark suits.  
All these benchmarks use OpenMP for parallelization and have known data races.  
The benchmarks cover OpenMP v$4.5$ pragmas comprehensively and contain race conditions due to common mistakes in OpenMP programming as listed earlier (Section~\ref{Se:Examples}).

DataRaceBench v1.2~\cite{liao2017dataracebench}, a \emph{seeded} OpenMP benchmark with known data race conditions, consists of $116$ microbenchmark kernels, out of which $59$ kernels have true data races, while the remaining $57$ kernels do not have any data races.
Out of these $59$ true race kernels, $44$ kernels have exactly one race condition.
The remaining $15$ have more than one race condition due to read and write operations to a scalar variable inside a loop.
Such cases have all the three types of dependences, namely read after write (RAW), write after read (WAR), and write after write (WAW) dependences and thus result in more than one race.
We have considered the location of the write operation for these cases and hence considered one TP per kernel.
DataRaceBench FORTRAN has $52$ kernels with true data races and $40$ kernels without any data races.
OmpSCR v$2.0$ is a benchmark suite for high performance computing using OpenMP v3.0 APIs.
The benchmark consists of C/C++ and FORTRAN kernels which demonstrate the usefulness and pitfalls of the parallel programming paradigm with both correct and incorrect parallelization strategies.
The kernels range from parallelization of simple loops with dependences to more complex parallel implementations of algorithms such as Mandelbrot set generator, Molecular Dynamics simulation, Pi ($\pi$) calculation, LU decomposition, Jacobi solver, fast Fourier transforms (FFT), and Quicksort.

\noindent\textbf{Performance Metrics Notations}. We define terminology used for performance metrics as follows:

\begin{itemize}
\item \textbf{True Positive (TP):}
If the evaluation tool correctly detects a data race present in the kernel, it is a True Positive test result. 
A higher number of true positives represents a better tool.
\item \textbf{True Negative (TN):}
If the benchmark does not contain a race and the tool declares it as race-free, then it is a true negative case. 
A higher number of true negatives represents a better tool.
\item \textbf{False Positives (FP):}
If the benchmark does not contain any race but the tool reports a race condition, then it is a false positive case.
A lower number of false positives are desirable.
\item \textbf{False Negatives (FN):} 
False Negative test result is obtained when the tool fails to detect a known race in the benchmark. 
These are the cases that are missed by the tool. 
A lower number of false negatives are desirable.
\end{itemize}

We consider the following statistical measures as performance metrics in our experiments.
\begin{itemize}
\item \textbf{Precision:} Precision is the measure of closeness of the outcomes of prediction.
Thus, a higher value of precision represents that the tool will more often than not identify a race condition when it exists. \\
\begin{minipage}{0.5\textwidth}
$Precision = \frac{TP}{TP\ +\ FP}$
\end{minipage}
\begin{minipage}{0.5\textwidth}
$Recall = \frac{TP}{TP\ +\ FN}$
\end{minipage}
\item \textbf{Recall:} Recall gives the total number of cases detected out of the maximum data races present. 
A higher recall value means that there are less chances that a data race is missed by the tool.
It is also called true positive rate (TPR).
\item \textbf{Accuracy:} Accuracy gives the chances of correct reports out of all the reports, as the name suggests. 
A higher value of accuracy is always desired and gives overall measure of the efficacy of the tool. \\
\begin{minipage}{0.5\textwidth}
$Accuracy = \frac{TP\ +\ TN}{TP\ +\ FP\ +\ TN\ +\ FN}$ 
\end{minipage}
\begin{minipage}{0.5\textwidth}
$F1\ Score = 2 * \frac{Precision\ *\ Recall}{Precision\ +\ Recall}$
\end{minipage}
\item \textbf{F1 Score:} The harmonic mean of precision and recall is called the F1 score.
An F1 score of $1$ can be achieved in the best case when both precision and recall are perfect.
The worst case F1 score is 0 when either precision or recall is 0.
\item \textbf{Diagnostic odds ratio (DOR):}  It is the ratio of the positive
likelihood ratio (LR$+$) to the negative likelihood ratio (LR$-$).\\
$DOR = \frac{LR+}{LR-}$
where, \\
\begin{minipage}{0.5\textwidth}
Positive Likelihood Ratio $(LR+) = \frac{TPR}{FPR}$ ,\\
Negative Likelihood Ratio $(LR-) = \frac{FNR}{TNR}$ ,\\
True Positive Rate $(TPR) = \frac{TP}{TP\ +\ FN}$ ,
\end{minipage}
\begin{minipage}{0.5\textwidth}
False Positive Rate $(FPR) = \frac{FP}{FP\ +\ TN}$ ,\\
False Negative Rate $(FNR) = \frac{FN}{FN\ +\ TP}$ and \\
True Negative Rare $(TNR) = \frac{TN}{TN\ +\ FP}$
\end{minipage}
\textbf{DOR} is the measure of the ratio of the odds of race detection being positive given that the test case has a data race, to the odds of race detection being positive given the test case does not have a race.
\end{itemize}

\noindent\textbf{System configuration}. We performed all our experiments on a system with two Intel Xeon E5-2697 v4 $@$ 2.30GHz processors, each having $18$ cores and $2$ threads per core, totalling $72$ threads and $128$GB of RAM.
The system runs 64 bit Ubuntu $18.04.2$ LTS server with Linux kernel version $4.15.0$-$48$-generic. 
\toolname is currently based on the LLVM/Polly version release $7.0.1$ 
and can be upgraded to the latest LLVM/Polly versions with minimal changes.

Similar to Liao et al.~\cite{liao2017dataracebench}, our experiments use two parameters: (i) the number of OpenMP threads  and (ii) the input size for variable length arrays.  The number of threads that we considered for the experiments are $\{3,36,45,72,90,180,256\}$. 
For the 16 variable length kernels, we considered 6 different array sizes as follows: $\{32,64,128,256,512,1024\}$.
With each particular set of parameters, we ran each of the $116$ kernels $5$ times.
Both the number of threads and array sizes can be found in prior studies~\cite{liao2017dataracebench,liao2017dataracebenchv12} and we have used the same for uniformity.
Since the dynamic tools depend on the execution order of the threads, multiple runs are required. 
The 16 kernels with variable length arrays were run $3360$ ( $16$ kernels $\times$ $7$ thread sizes $\times$ $6$ array sizes $\times$ $5$ runs) times in total.
The remaining $100$ kernels were run $3500$ ($100$ kernels $\times$ $7$ thread sizes $\times$ $5$ runs ) times in total.
For all experiments, we used a timeout of $600$ seconds for compilation as well as execution separately.

\subsection{Experimental Results\label{Se:exresults}}
\subsubsection{DataRaceBench $1.2$}
\begin{table}[t]
\caption{Maximum number of Races reported by different tools in DataRaceBench 1.2}
\centering
\begin{tabular}{ |l|cc|cc|r| }
 \hline
 \cline{1-6}
 \multirow{2}{*}{Tools} & \multicolumn{2}{c|}{Race: Yes} & \multicolumn{2}{c|}{Race: No} &
 \multirow{2}{*}{Coverage/116}  \\ 
 \cline{2-5} & TP & FN & TN & FP & \\
 \hline
 \tool{Helgrind} & 56  & 3 & 2 & 55 & 116 \\
 \hline
 \tool{Valgrind DRD} & 56 & 3 & 26 & 31 & 116 \\
 \hline
 \tool{TSan-LLVM} & \textbf{57} & 2 & 2 & 55 & 116 \\
 \hline
 \tool{Archer} & 56 & 3 & 2 & 55 & 116 \\
 \hline
 \tool{SWORD} & 47 & 4 & 24 & 4 & 79 \\
 \hline
 \hline
\toolname & 48 & 2 & \textbf{36} & 5 & 91 \\
 \hline
\end{tabular}
\label{tbl:comparison}
\end{table}

\tableref{comparison} provides comparison in terms of the number of races detected in DataRaceBench v1.2.
Column 1 indicates the name of the tool. The column with titles ``Race:Yes'' and ``Race:No'' indicates if the benchmark had a race or not. 
The subcolumns ``TP'' and ``FN'' denote whether the tool was able to find the race or not when the benchmark had a race. 
Similarly, the subcolumns ``FP'' and ``TN'' denote if the tool erroneously reported a race or reported the absence of a race when the benchmark did not have a race. 
Dynamic analysis tools are run multiple times, and, if the tool reports a race in any of the runs, then, it is considered that the tool will classify the benchmark as having a race.

\toolname could analyze $91$ out of $116$ ($78.45\%$) kernels from the benchmark, and could detect $48$ True Positives (TP) and $36$ True Negatives (TN).
As the analysis of our tool is conservative, it also produces $5$ False Positives (FP).
Moreover, it had $2$ False Negatives (FN) because of inter SCoP races which is a limitation of the current version of LLOV.

Due to the sound static analysis that \toolname implements, it could also prove that $36$ of the kernels are \emph{data race free}.
\toolname is unique in this regard, other tools \textit{are not able to make} such a claim.
\toolname will report one of the following three cases: \emph{data race detected} when \toolname detects a race, 
\emph{data race free} when \toolname can statically prove that the parallel segment of the input program does not have dependences due to shared memory accesses, 
and finally, \emph{region not analyzed} when \toolname cannot analyze the input program.
In addition, \toolname also reports if  an OpenMP pragma is not supported (refer to \tableref{pragmas}). Due to these reasons, as of now \toolname does not provide complete coverage on DataRaceBench v1.2. \tool{SWORD} provides even lesser coverage on DataRaceBench v1.2 due to its compilation related issues.

\tableref{metrics} shows performance of the tool on various performance metrics defined earlier in the section.
From \tableref{metrics} it appears that \toolname's performance is the best followed by \tool{SWORD} in second in terms of precision, accuracy, F1 score and DOR.
Since both \tool{SWORD} and \tool{LLOV} do not have complete coverage, a more appropriate strategy would be to compare all the tools on only those benchmarks which they are able to handle/cover.

\vspace{-2em}
\begin{table}[H]
\caption{Precision, Recall and Accuracy of the tools on DataRaceBench 1.2}
\begin{tabular}{ |l|c|c|c|c|c| }
\hline
Tools & Precision & Recall & Accuracy & F1 Score & Diagnostic odds ratio \\
\hline
\tool{Helgrind} & 0.50 & 0.95	& 0.50 & 0.66 & 0.68 \\
\hline
\tool{Valgrind DRD} & 0.64 & 0.95	& 0.71 & 0.77 & 15.66 \\
\hline
\tool{TSan-LLVM} & 0.51 & \textbf{0.97} & 0.51 & 0.67 & 1.04 \\
\hline
\tool{Archer} & 0.50 & 0.95 & 0.50 & 0.66 & 0.68 \\
\hline
\tool{SWORD} & \textbf{0.92} & 0.92 & 0.90 & 0.92 & 70.50 \\	\hline
\hline
\toolname & 0.91 & 0.96 & \textbf{0.92} & \textbf{0.93} & \textbf{172.80} \\
 \hline
\end{tabular}
\label{tbl:metrics}
\end{table}

\vspace{-2em}
\begin{table}[H]
\caption{Maximum number of Races reported by different tools in common 61 kernels of DataRaceBench 1.2}
\centering
\begin{tabular}{ |l|cc|cc|r| }
 \hline
 \cline{1-6}
 \multirow{2}{*}{Tools} & \multicolumn{2}{c|}{Race: Yes} & \multicolumn{2}{c|}{Race: No} &
 \multirow{2}{*}{Coverage/61}  \\ 
 \cline{2-5} & TP & FN & TN & FP & \\
 \hline
 \tool{Helgrind} & 42  & 1 & 2 & 16 & 61 \\
 \hline
 \tool{Valgrind DRD} & 42 & 1 & 12 & 6 & 61 \\
 \hline
 \tool{TSan-LLVM} & 42 & 1 & 2 & 16 & 61 \\
 \hline
 \tool{Archer} & 42 & 1 & 2 & 16 & 61 \\
 \hline
 \tool{SWORD} & 42 & 1 & 17 & 1 & 61 \\
 \hline
 \hline
\toolname & 42 & 1 & 16 & 2 & 61 \\
\hline
\end{tabular}
\label{tbl:comparison-common}
\end{table}

\vspace{-2em}
\begin{table}[H]
\caption{Precision, Recall and Accuracy of the tools on common 61 kernels of DataRaceBench 1.2}
\resizebox{0.9\columnwidth}{!}{
\begin{tabular}{ |l|c|c|c|c|c| }
\hline
Tools & Precision & Recall & Accuracy & F1 Score & Diagnostic odds ratio \\
\hline
\tool{Helgrind} & 0.72 & \textbf{0.98}	& 0.72 & 0.83 & 5.25 \\
\hline
\tool{Valgrind DRD} & 0.88 & \textbf{0.98}	& 0.89 & 0.92 & 84.00 \\
\hline
\tool{TSan-LLVM} & 0.72 & \textbf{0.98} & 0.72 & 0.83 & 5.25 \\
\hline
\tool{Archer}  & 0.72 & \textbf{0.98} & 0.72 & 0.83 & 5.25 \\
\hline
\tool{SWORD} & \textbf{0.98} & \textbf{0.98} & \textbf{0.97} & \textbf{0.98} & \textbf{714.00} \\	\hline
\hline
\toolname & 0.95 & \textbf{0.98} & 0.95 & 0.97 & 336.00 \\
 \hline
\end{tabular}
}
\label{tbl:metrics-common}
\end{table}

\tableref{comparison-common} shows how tools classify benchmarks with respect to data race on $61$ benchmarks that all the tools are able to handle/cover.
\tableref{metrics-common} provides a comparison of the tools on $61$ benchmarks on various performance metrics. 
It is indeed the case that on the benchmarks that \tool{SWORD} is able to handle, it achieves the highest precision, accuracy, recall, F1 score and diagnostic odds ratio. 
\tool{LLOV} is a close second with respect to precision, accuracy and F1 score. 
One must keep in mind that both \tool{SWORD} and \tool{LLOV} may gain advantage in terms of these metrics because of lesser coverage.  A crucial point to note is that while \tool{SWORD} crashes on many benchmarks, \tool{LLOV} provides graceful reporting and exit on benchmarks it is not able to cover, providing a better user experience.

\begin{figure}[H]
\subfloat[Time taken on all 116 kernels]{%
\begin{tikzpicture}
\centering
\begin{axis}[
  scale = 0.5,
  bar width=5pt,
  ylabel=Execution time in Seconds (log scale),
  xlabel=Number of threads,
  ymin=1, ymax=25000,
  ymode=log,
  log basis y={10},
  log ticks with fixed point,
  scale only axis,
  scaled x ticks=manual,
  axis x line* =none, 
  axis y line* =none, 
  symbolic x coords={0, 3,36,45,72,90,180,256, 300},
  major x tick style = {opacity=0},
  minor x tick num = 1,
  minor y tick num = 10,
  minor tick length=2ex,
  every node near coord/.append style={
    anchor=east,
    rotate=90
  },
  xtick={data},
  xtick distance = 5,
  xtick pos=left,
  extra y ticks={44.1},
  extra y tick style={
	ytick align=outside,
	tick style={thick,teal,},
	yticklabel style={
		font=\small,
		color=teal,
        /pgf/number format/.cd,
        fixed,
        fixed zerofill,
        precision=2,
	},
  },
  ytick pos=left,
  scaled x ticks = true,
  scaled y ticks = false,
  legend columns = 2,
  legend style={ 
    column sep=1ex,
    at={(0.55,0.35)}, 
	anchor=north,
	scale = 0.2,
	font=\scriptsize,
  },
  title=Lower time is better,
]
\addplot+[
  color=red,
  fill=none,
 ]
table {./Data/Archer_TT.txt};
\addlegendentry{Archer}
\addplot+[
  color=green,
  fill=none,
 ]
table {./Data/Helgrind_TT.txt};
\addlegendentry{Helgrind}
\addplot+[
  color=blue,
  fill=none,
 ]
table {./Data/Sword_TT.txt};
\addlegendentry{\tool{SWORD}}
\addplot+[
  color=orange,
  fill=none,
]
table {./Data/Tsan_TT.txt};
\addlegendentry{\tool{TSan}}
\addplot+[
  color=gray,
  fill=none,
 ]
table {./Data/DRD_TT.txt};
\addlegendentry{DRD}
\addplot+ [teal, very thick, mark=none, line legend, sharp plot,update limits=false,]
coordinates { (0,44.1) (300,44.1)}
node [above] at (0,44.1){};
\addlegendentry{\toolname}
\end{axis}
\label{fig:drb-116}
\end{tikzpicture}
}
\hspace*{5mm}
\subfloat[Time taken on common 61 kernels]{%
\begin{tikzpicture}
\begin{axis}[
  scale = 0.5,
  bar width=5pt,
  ylabel=Execution time in Seconds (log scale),
  xlabel=Number of threads,
  ymin=1, ymax=3000,
  ymode=log,
  log basis y={10},
  log ticks with fixed point,
  scale only axis,
  scaled x ticks=manual,
  axis x line* =none, 
  axis y line* =none, 
  symbolic x coords={0, 3,36,45,72,90,180,256, 300},
  major x tick style = {opacity=0},
  minor x tick num = 1,
  minor y tick num = 10,
  minor tick length=2ex,
  every node near coord/.append style={
    anchor=east,
    rotate=90
  },
  xtick={data},
  xtick distance = 5,
  xtick pos=left,
  extra y ticks={4.3},
  extra y tick style={
	ytick align=outside,
	tick style={thick,teal,},
	yticklabel style={
		font=\small,
		color=teal,
	},
  },
  ytick pos=left,
  scaled x ticks = true,
  scaled y ticks = false,
  legend pos=north west,
  legend columns = 2,
  legend style={ 
    column sep=1ex,
    at={(0.6,0.5)},
	anchor=north,
	font=\scriptsize,
	scale = 0.2,
  },
  title=Lower time is better,
]
\addplot+[
  color=red,
  fill=none,
 ]
table {./Data/Archer_drb_66.txt};
\addlegendentry{Archer}
\addplot+[
  color=green,
  fill=none,
 ]
table {./Data/Helgrind_drb_66.txt};
\addlegendentry{Helgrind}
\addplot+[
  color=blue,
  fill=none,
 ]
table {./Data/Sword_drb_66.txt};
\addlegendentry{\tool{SWORD}}
\addplot+[
  color=orange,
  fill=none,
]
table {./Data/Tsan_drb_66.txt};
\addlegendentry{\tool{TSan}}
\addplot+[
  color=gray,
  fill=none,
 ]
table {./Data/DRD_drb_66.txt};
\addlegendentry{DRD}
\addplot+ [teal, very thick, mark=none, line legend, sharp plot,update limits=false,]
    coordinates { (0,4.3) (300,4.3)}
    node [above] at (0,4.3){};
\addlegendentry{
\toolname}
\end{axis}
\label{fig:drb-common}
\end{tikzpicture}
}
\caption{DataRaceBench v$1.2$ total time taken on logarithmic scale}
\label{fig:drb-two}
\end{figure}
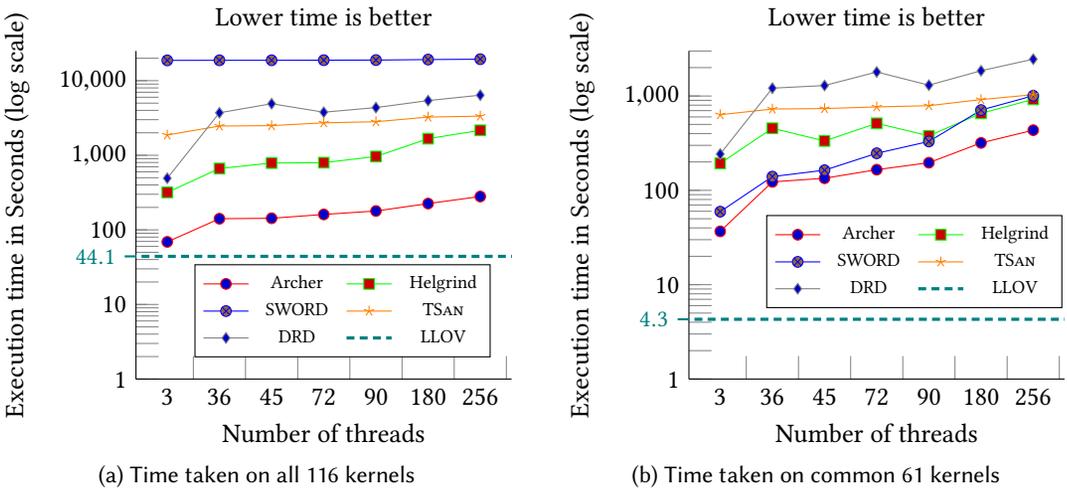

Though \toolname does not come out on top on various metrics such as coverage, precision etc., it completely outshines other tools in terms of runtime.
\figref{drb-116} shows the performance of the tool with respect to the run time.
Since dynamic tools run benchmarks multiple times we report the average time taken for each benchmark, and the total time is the sum of these averages.
In \figref{drb-116}, $y$-axis represents total time taken in seconds by a tool on a logarithmic scale.
Time taken by \toolname to analyze all $116$ kernels is a mere $44.1$ seconds.
On the other hand, other tools take orders of magnitude more time as compared to \toolname.
The reason for \tool{SWORD} performing the worst when all $116$ benchmarks are considered is because the compilation process itself times out for several benchmarks.
Timeout value of $600$ seconds is used for each kernel for both compilation and execution separately.
For \toolname the only time required is the time to compile as it does its analysis at compile time.
In addition, as \toolname is able to detect the cases it can not analyze, the exits for such programs are graceful.
The power of static analysis in \toolname is particularly evident in \figref{drb-two} as the time taken remains constant irrespective of the number of threads.

The Polyhedral framework is known for large compile-time overheads because it relies on computationally expensive algorithms for dependence analysis, scheduling and code-generation.
These algorithms could be exponential, or polynomials of high-degree in complexity in the number of dimensions in the loop nest~\cite{feautrier1988parametric, Upadrasta-Cohen:2013, acharya2018autotune, Bas04b, verdoolaege2010isl}.
However, very few programs in the real world have very large loop-depths.

In DataRaceBench v$1.2$, there are six tiled and parallel versions of PolyBench/C $3.2$ kernels.
The tiled version of matrix multiplication kernel DRB042-3mm-tile-no has $408$ OpenMP parallel loops and contributes to around $69.21\%$ of total time taken by \toolname for all the 116 kernels.
Although such kernels are not very common in real world scenarios, as they are  generated by polyhedral tools such as PLuTo~\cite{Bondhugula:2008:PAP:1375581.1375595}, verification of code generated by such automatic tools remains a challenge.

\figref{drb-common} shows the runtime performance of the tools on $61$ benchmarks that all the tools are able to cover. 
On this subset of benchmarks, \toolname outperforms all the other tools  by orders of magnitude. 
\tool{Archer} is second in performance and \tool{SWORD} comes third. 

\vspace{-2em}

\subsubsection{DataRaceBench $1.2$ FORTRAN}
Since \toolname is based on LLVM-IR, it is language independent.   To demonstrate this, we reimplemented DataRaceBench $1.2$ in FORTRAN $95$~\cite{drbForturl} rewriting $92$ of the $116$ DataRaceBench v$1.2$ kernels in FORTRAN.
The other kernels, such as $\{41,42,43,44,55,56\}$, are Polybench kernels that were tiled and/or parallelized by the POCC~\cite{poccurl} polyhedral tool, and are not amenable to easy re-writing in FORTRAN.

The kernel in \lstref{datasharingclause_F} (~\sectref{Examples}) is from DataRaceBench $1.2$ FORTRAN which has a data race because of the write to \TT{shared} variable \TT{tmp}~\lstline{tmpwrite} and the read from \TT{tmp}~\lstline{tmpread}.
The race in this example can be avoided by explicitly stating that \TT{tmp} is a private variable for each thread using the \TT{private} clause.
\toolname could detect such races because of missing data sharing clauses, such as \TT{private}, \TT{reduction}, \TT{firstprivate}, and \TT{lastprivate}.

To verify these FORTRAN kernels, we used LLVM FORTRAN frontend Flang~\cite{flang} which is under active development and has officially been accepted in the year 2019 as a sub-project under the LLVM Compiler Infrastructure umbrella project.  We used Flang version $7.1.0$ (git commit hash cb42a171) to generate LLVM-IR from FORTRAN source code and ran \toolname on the generated IR.

\vspace{-2em}
\begin{table}[H]
\caption{Maximum number of Races reported by different tools in DataRaceBench FORTRAN}
\begin{tabular}{ |l|cc|cc|r| }
 \hline
 \cline{1-6}
 \multirow{2}{*}{Tools} & \multicolumn{2}{c|}{Race: Yes} & \multicolumn{2}{c|}{Race: No} &
 \multirow{2}{*}{Coverage/92}  \\ 
 \cline{2-5} & TP & FN & TN & FP & \\
 \hline
 \tool{Helgrind} & \textbf{46}  & 6 & 4 & 36 & 92 \\
 \hline
 \tool{Valgrind DRD} & 45 & 7 & \textbf{21} & 19 & 92 \\
 \hline
 \hline
\toolname & 36 & 7 & 19 & 5 & 67 \\
 \hline
\end{tabular}
\label{tbl:drb_fortran-comparison}
\end{table}

\vspace{-1em}

Initial experiments show that our analysis is able to detect race conditions in OpenMP kernels of DataRaceBench FORTRAN. To the best of our knowledge, \toolname is the \EM{only static tool} to be able to detect races in OpenMP programs written in FORTRAN.
\tableref{drb_fortran-comparison} shows that \toolname could detect $36$ True Positives (TP) and also confirm that $19$ kernels are data race free (TN).
\toolname also produced $5$ False Positives (FP) along with $7$ False Negatives (FN).

Kernels $\{72, 78, 79, 94, 112\}$ are not analyzed by \toolname because the current version of Flang does not support the corresponding OpenMP directives.  This explains the difference in the numbers from DataRaceBench v$1.2$ (\tableref{comparison}) and DataRaceBench FORTRAN (\tableref{drb_fortran-comparison}).
Flang produced segmentation faults for the kernels $\{84, 85\}$ having OpenMP \TT{threadprivate} variables.
As Flang is in active development, we believe that more OpenMP directives will be supported in the upcoming releases. 
Another challenge we faced is detecting polyhedral SCoPs in Flang generated IR using Polly~\cite{polly}. 
Even for functionally equivalent source codes, the IR generated by Clang and Flang differs considerably.
So, the native implementations of analyses and optimization passes in LLVM need to be modified to have uniform results on LLVM-IR generated by Clang and Flang.

\subsubsection{OmpSCR v2.0}

We also evaluated all the tools listed in \tableref{Tools} on OmpSCR v$2.0$~\cite{ompSCR,dorta2005ompSCR} kernels.
For the dynamic tools, we used default program parameters provided with argument \emph{-test}.
We have done minor modifications in OmpSCR v2.0~\cite{ompSCR} to compile it on the latest operating systems with updated system calls, 
and created scripts to run and test all the data race checkers.
The updated version of OmpSCR can be found at~\cite{ompSCRgit}.

We manually verified the OmpSCR v$2.0$ benchmark suite. \tableref{ompscr_3} divides the benchmarks
into three categories: 
1) Manually verified kernels with data races, 
2) Manually verified race-free kernels,
and 3) Unverified kernels.
Out of $25$ kernels, $11$ remained unverified due to various complexities such as recursive calls using OpenMP pragmas.
Every cell in~\tableref{ompscr_3} denotes how many different regions are reported as containing a race by a tool.
Every verified kernel having a race contains only one parallel region containing races. 
All the races reported by a tool belonging to a single parallel region is counted as only one.
The reason for combining all the races in a region is because otherwise, the number of races reported for dynamic
tools becomes quite high (e.g., several races reported for a single array).
Some of the programs contain multiple parallel regions. 
If a tool reports races in two distinct regions, then we count the tool as reporting $2$ races. 
Since every verified kernel with race has only
one parallel region having a race, races reported in other race-free regions are counted towards \EM{false positives}.

As shown in \tableref{ompscr_3}, \toolname is able to detect true data races in c\_loopA and c\_loopB kernels.
\toolname produced a false negative for Jacobi03 kernel due to the presence of a dependence across two SCoPs, which is a limitation of the current version of \toolname. 
Our tool produced false positives for all three Jacobi kernels due to conservative Mod/Ref analysis in LLVM.
All the three Jacobi kernels in OmpSCR use a one dimensional array which is passed to the kernel as a pointer.
Also, the programs are written with the arrays accessed in the column-major order, which means that their array subscripts are incorrectly computed in C/C++.
Because of these reasons, the three Jacobi kernels are not modelled by Polly even though Jacobi is a standard polyhedral kernel.

\vspace{-2em}
\begin{table}[H]
\caption{Number of Races detected in OmpSCR v2.0 (CT is Compilation Timeout, NA for Not Analyzed)}
\small
\resizebox{0.9\columnwidth}{!}{
\begin{tabular}{ |l|c|c|c|c|c|c| }
\hline
\textbf{Kernel}     & \textbf{\toolname} & \textbf{\tool{Helgrind}} & \textbf{\tool{DRD}} & \textbf{\tool{TSan-LLVM}} & \textbf{\tool{Archer}} &  \textbf{\tool{SWORD} } \\
\hline
\multicolumn{7}{|c|}{Manually verified kernels with data races} \\
\hline
c\_loopA.badSolution         & 1 & 1 & 1  & 1  & 1  & 1       \\
c\_loopA.solution2           & NA & 1 & 1  & 1  & 1  & 0       \\
c\_loopA.solution3           & 1 & 1 & 1  & 1  & 1  & 0       \\
c\_loopB.badSolution1        & 1 & 1 & 1  & 1  & 1  & 1       \\
c\_loopB.badSolution2        & 1 & 1 & 1  & 1  & 1  & 1       \\
c\_loopB.pipelineSolution    & NA & 1 & 1  & 1  & 1  & 0       \\
c\_lu                        & NA & 1 & 1  & 1  & 1  & 0       \\
c\_jacobi03                  & 1 & 1 & 1  & 0  & 0  & CT      \\
\hline
\multicolumn{7}{|c|}{Manually verified race free kernels} \\
\hline
c\_loopA.solution1           & 0 & 2 & 1  & 2  & 1  & 0       \\
c\_md                        & 1 & 2 & 2  & 2  & 1  & CT      \\
c\_mandel                    & NA & 1 & 0  & 1  & 1  & 0       \\
c\_pi                        & 0 & 1 & 0  & 1  & 1  & 0       \\
c\_jacobi01                  & 2 & 2 & 1  & 0  & 0  & CT      \\
c\_jacobi02                  & 1 & 1 & 1  & 0  & 0  & CT      \\
\hline
\multicolumn{7}{|c|}{Unverified kernels} \\
\hline
c\_fft                       & 0 & 1 & 1  & 1  & 1  & CT      \\
c\_fft6                      & 2 & 1 & 0  & 1  & 1  & CT      \\
c\_qsort                     & 0 & 1 & 1  & 1  & 1  & CT      \\
c\_GraphSearch                 & 0 & 0 & 0  & 0  & 0  & 0      \\
cpp\_qsomp1                     & 0 & 0 & 0  & 0  & 0  & 0      \\
cpp\_qsomp2                     & 0 & 0 & 0  & 0  & 0  & 0      \\
cpp\_qsomp3                     & 0 & 0 & 0  & 0  & 0  & 0      \\
cpp\_qsomp4                     & 0 & 0 & 0  & 0  & 0  & 0      \\
cpp\_qsomp5                     & 1 & 0 & 0  & 0  & 0  & 0      \\
cpp\_qsomp6                     & 0 & 0 & 0  & 0  & 0  & 0      \\
cpp\_qsomp7                     & 0 & 0 & 0  & 0  & 0  & 0      \\
\hline
\end{tabular}
}
\label{tbl:ompscr_3}
\end{table}

\vspace{-2em}
\begin{table}[H]
\caption{Comparison of different tools on OmpSCR v2.0}
\begin{tabular}{ |l|cc|cc|r| }
 \hline
 \cline{1-6}
 \multirow{2}{*}{Tools} & \multicolumn{2}{|c|}{Race: Yes} & \multicolumn{2}{|c|}{Race: No} &
 \multirow{2}{*}{Coverage/14}  \\ 
 \cline{2-5} & TP & FN & TN & FP & \\
 \hline
 \tool{Helgrind} & 8 & 0 & 0 & 9 & 14 \\
 \hline
 \tool{Valgrind DRD} & 8 & 0 & 2 & 5 & 14 \\
 \hline
 \tool{TSan-LLVM} & 7 & 1 & 2 & 6 & 14 \\
 \hline
 \tool{Archer} & 7 & 1 & 2 & 4 & 14 \\
 \hline
 \tool{SWORD} & 3 & 4 & 3 & 0 & 10 \\
 \hline
 \hline
\toolname & 4 & 1 & 2 & 5 & 10 \\
 \hline
\end{tabular}
\label{tbl:comparison_ompscr}
\end{table}

\vspace{-1em}
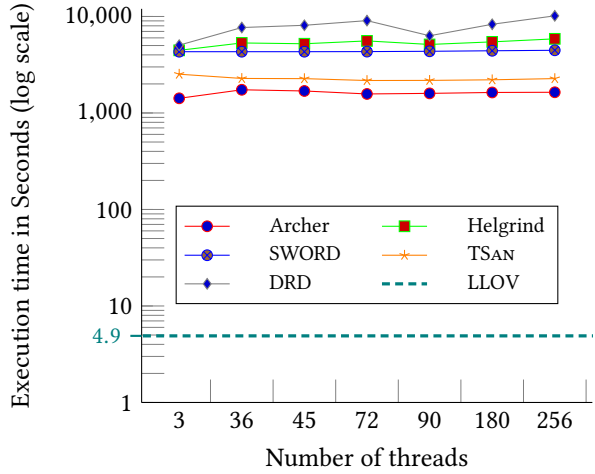
\begin{figure}[H]
\begin{tikzpicture}
\begin{axis}[
  scale=.6,
  bar width=5pt,
  ylabel=Execution time in Seconds (log scale),
  xlabel=Number of threads,
  ymin=1, ymax=12000,
  ymode=log,
  log basis y={10},
  log ticks with fixed point,
  scale only axis,
  scaled x ticks=manual,
  axis x line* =none, 
  axis y line* =none, 
  symbolic x coords={0, 3,36,45,72,90,180,256, 300},
  major x tick style = {opacity=0},
  minor x tick num = 1,
  minor y tick num = 10,
  minor tick length=2ex,
  every node near coord/.append style={
    anchor=east,
    rotate=90
  },
  xtick={data},
  xtick distance = 5,
  xtick pos=left,
  extra y ticks={4.9},
  extra y tick style={
	ytick align=outside,
	tick style={thick,teal,},
	yticklabel style={
		font=\small,
		color=teal,
	},
  },
  ytick pos=left,
  scaled x ticks = true,
  scaled y ticks = false,
  legend pos=outer north east,
  legend plot pos=left,
  legend cell align=left, 
  legend columns = 2,
  legend style={ 
    font=\footnotesize,
    column sep=3ex,
    at={(0.5,0.5)},
	anchor=north,
  },
]
\addplot+[
  color=red,
 ]
table {./Data/Archer_ompscr.txt};
\addlegendentry{
Archer}
\addplot+[
  color=green,
 ]
table {./Data/Helgrind_ompscr.txt};
\addlegendentry{Helgrind}
\addplot+[
  color=blue,
 ]
table {./Data/Sword_ompscr.txt};
\addlegendentry{\tool{SWORD}}
\addplot+[
  color=orange,
]
table {./Data/TSan_ompscr.txt};
\addlegendentry{\tool{TSan}}
\addplot+[
  color=gray,
 ]
table {./Data/DRD_ompscr.txt};
\addlegendentry{DRD}
\addplot+ [teal, very thick, mark=none, line legend, sharp plot,update limits=false,]
coordinates { (0,4.9) (300,4.9)}
node [above] at (0,4.9){};
\addlegendentry{
\toolname}
\end{axis}
\end{tikzpicture}
\caption{OmpSCR v$2.0$ total execution time by different tools on logarithmic scale}
\label{fig:OmpSCRlogline}
\end{figure}

All the false positives flagged by \toolname are because of \TT{shared} double pointer variables.
In addition, \tool{SWORD} ends up with compiler timeout (CT) for kernels such as Molecular Dynamics, Quicksort (c\_qsort), FFT, and Jacobi.
The time taken to detect races in all the kernels by the tools is shown in \figref{OmpSCRlogline}.
It can be seen that \toolname completes its analysis for the entire benchmark in just \emph{4.9 seconds}, while the other state-of-the-art tools take orders of magnitude longer.
\section{Conclusions and Future work}
\label{Se:futurework}

In this paper, we present \toolname, a language agnostic, OpenMP aware, static analysis based data race detection tool which  is developed on top of the LLVM compiler framework. 
As \toolname operates at LLVM-IR level, it can support a multitude of programming languages supported by the LLVM infrastructure. 
We successfully demonstrate the language agnostic nature of \toolname by performing data race checks on a standard set of benchmarks written in C, C++, and FORTRAN.

Our experiments show that \toolname performs reasonably well in terms of precision and accuracy while being \textit{most performant} with respect to other tools by a large margin. Though at present, \toolname supports only some of the pragmas offered by OpenMP, it gracefully exits on input programs that contain pragmas that it is unable to handle. 
We would like to further enrich \toolname by adding support for various OpenMP pragmas that are not supported at present.
Many such pragmas offer an engineering challenge as the structural information is not available at LLVM-IR level, and such information has to be reconstructed from the IR.

Our tool is primarily based on the polyhedral compilation framework, Polly.
The use of approximate dependence analysis~\cite{llvm-lai} readily available in LLVM may further increase the capability and scalability of \toolname. 
We also plan to extend the support for dynamic control flow and irregular accesses using the extended polyhedral framework~\cite{collard1995fada, belaoucha2010fadalib, benabderrahmane2010widely}.
We plan to use May-Happen-in-Parallel (MHP) analysis to provide coverage for OpenMP tasks, synchronizations, sections constructs, and overcome the current limitation of FN cases for dependences across SCoPs.

The tool and other relevant material are available at: \url{http://compilers.cse.iith.ac.in/projects/llov}.
\newpage
\begin{acks}

We thank Tobias Grosser, Johannes Doerfert, Michael Kruse for their help with the initial version of Polly as an analysis pass, which we extended for this work. We also thank Govindarajan Ramaswamy and V Krishna Nandivada for their feedback on this work.
We would like to thank the anonymous reviewers of ACM TACO for their insightful comments which helped in improving the paper.

This work is partially supported by a fellowship under Visvesvaraya PhD Scheme from MeitY, India (grant PhD-MLA/04(02)/2015-16), an Early Career Research award from SERB, DST, India (grant ECR/2017/001126), a Visvesvaraya Young Faculty Research Fellowship from MeitY (MeitY-PHD-1149), an NSM research grant (sanction number MeitY/R\&D/HPC/2(1)/2014), a faculty research grant from AMD, and an Army Research Office grant (W911NF).

\end{acks}

\bibliographystyle{ACM-Reference-Format}
\bibliography{references}

\newpage
\newpage
\begin{appendices}
\section{Additional tables for DataRaceBench}

In the tables 3 and 5 on page 16 and 17 respectively\Omit{\tableref{comparison} and \tableref{comparison-common}}, we considered a single outcome of TP/FP/TN/FN for each kernel irrespective of the number of races detected by the particular tool.
  Hence, multiple races reported for a kernel with true races contributed to only one TP. Similarly, multiple races reported for a kernel that is race free contributed to only one FP.
  In the above tables, we also did not take into account the FP cases present within the kernels with true races.

  In the following tables, we provide the \textit{approximate count} of the total races reported by each tool for DataRaceBench kernels, including the FP cases reported for the kernels with true races.
  It is pertinent to remember that dynamic tools report \textit{the same race multiple times} across different executions with/without different program parameters. Here we count them as a \textit{single race}.
  We did this tabulation after \textit{manually} going through the logs generated by the tools.

  \tableref{drb-with-fp-in-tr} presents a comparison of the tools on DataRaceBench on the total number of races detected.
  Here we have included the races reported for \TT{pthread} mutex lock access by \tool{TSan-LLVM} and \tool{Archer}, and races detected at \TT{pthread} creation time by \tool{Helgrind} and \tool{Valgrind DRD}.
  The subcolumn ``FP'' under the column ``Race: Yes'' represents the number of FPs reported by a tool in the kernels with true races.
  Even though there are only $59$ kernels with true races, the number of FPs reported in these kernels by 
  the tools \tool{Helgrind}, \tool{Valgrind DRD} and \tool{TSan-LLVM} are more than $59$ because the numbers represent the approximate count of the total reported races.
  A significant number of these races reported by \tool{TSan-LLVM}, \tool{Helgrind}, and \tool{Valgrind DRD} are FPs.
  The performance of \tool{Archer} in reducing the FPs reported by \tool{TSan-LLVM} is commendable.
  Most of the FP races ($16$ out of $21$ in total) reported by \toolname are for a single kernel (DRB041), a parallelized version of matrix multiplication, which is generated by an automatic parallelization tool.

  \tableref{metrics-full-FP-in-TR-no-mutexlock} presents the various performance metrics for the numbers corresponding to \tableref{drb-with-fp-in-tr}.
\vspace{-2em}
\begin{table}[H]
  \caption{Maximum number of races reported by different tools in DataRaceBench 1.2 including FP in True Race kernels with the races because of \TT{pthread} mutex lock accesses}
\centering
\resizebox{0.6\columnwidth}{!}{
\begin{tabular}{ |l|ccc|cc|r| }
 \hline
 \cline{1-7}
 \multirow{2}{*}{Tools} & \multicolumn{3}{c|}{Race: Yes} & \multicolumn{2}{c|}{Race: No} &
 \multirow{2}{*}{Coverage/116}  \\
 \cline{2-6} & TP & FN & FP & TN & FP & \\
 \hline
 \tool{Helgrind} & 56  & 3 & 115 & 2 & 179 & 116 \\
 \hline
 \tool{Valgrind DRD} & 56 & 3 & 116 & 26 & 147 & 116 \\
 \hline
 \tool{TSan-LLVM} & \textbf{57} & 2 & 78 & 2 & 200 & 116 \\
 \hline
 \tool{Archer} & 56 & 3 & 56 & 2 & 59 & 116 \\
 \hline
 \tool{SWORD} & 47 & 4 & 6 & 24 & 4 & 79 \\
 \hline
 \hline
\toolname & 48 & 2 & 0 & \textbf{36} & 21 & 91 \\
 \hline
\end{tabular}
}
\label{tbl:drb-with-fp-in-tr}
\end{table}

\vspace{-2em}
\begin{table}[H]
  \caption{Precision, Recall and Accuracy of the tools on DataRaceBench 1.2 including FP from True Race kernels with races due to \TT{pthread} mutex lock accesses}
\resizebox{0.8\columnwidth}{!}{
\begin{tabular}{ |l|c|c|c|c|c| }
\hline
Tools & Precision & Recall & Accuracy & F1 Score & Diagnostic odds ratio \\
\hline
\tool{Helgrind} & 0.16 & 0.95 & 0.16 & 0.27 & 0.13 \\
\hline
\tool{Valgrind DRD} & 0.18 & 0.95 & 0.24 & 0.30 & 1.85 \\
\hline
\tool{TSan-LLVM} & 0.17 & \textbf{0.97} & 0.17 & 0.29 & 0.21 \\
\hline
\tool{Archer} & 0.33 & 0.95 & 0.33 & 0.49 & 0.32 \\
\hline
\tool{SWORD} & \textbf{0.82} & 0.92 & \textbf{0.84} & \textbf{0.87} & 28.20 \\	\hline
\hline
\toolname & 0.70 & 0.96 & 0.79 & 0.81 & \textbf{41.14} \\
 \hline
\end{tabular}
}
\label{tbl:metrics-full-FP-in-TR-no-mutexlock}
\end{table}

\tableref{drb-with-fp-in-tr-no-mutexlock} presents a comparison of the tools on DataRaceBench on total number of races detected excluding the races reported for \TT{pthread} mutex lock access by 
\tool{TSan-LLVM} and \tool{Archer}, and races detected at \TT{pthread} creation time by \tool{Helgrind} and \tool{Valgrind DRD}.
It can be seen that the FPs reported by the tool \textit{reduced significantly} when compared to \tableref{drb-with-fp-in-tr}.\\
  \tableref{metrics-full-FP-in-TR} presents the various performance metrics corresponding to \tableref{drb-with-fp-in-tr-no-mutexlock}.
\vspace{-2em}
\begin{table}[H]
  \caption{Maximum number of races reported by different tools in DataRaceBench 1.2 including FP in True Race kernels excluding the races because of \TT{pthread} mutex lock accesses}
\centering
\resizebox{0.6\columnwidth}{!}{
\begin{tabular}{ |l|ccc|cc|r| }
 \hline
 \cline{1-7}
 \multirow{2}{*}{Tools} & \multicolumn{3}{c|}{Race: Yes} & \multicolumn{2}{c|}{Race: No} &
 \multirow{2}{*}{Coverage/116}  \\
 \cline{2-6} & TP & FN & FP & TN & FP & \\
 \hline
 \tool{Helgrind} & 56 & 3 & 58 & 27 & 124 & 116 \\
 \hline
 \tool{Valgrind DRD} & 56 & 3 & 59 & 27 & 121 & 116 \\
 \hline
 \tool{TSan-LLVM} & \textbf{57} & 2 & 20 & 25 & 139 & 116 \\
 \hline
 \tool{Archer} & 56 & 3 & 0 & \textbf{53} & 4 & 116 \\
 \hline
 \tool{SWORD} & 47 & 4 & 6 & 24 & 4 & 79 \\
 \hline
 \hline
\toolname & 48 & 2 & 0 & 36 & 21 & 91 \\
 \hline
\end{tabular}
}
\label{tbl:drb-with-fp-in-tr-no-mutexlock}
\end{table}

\vspace{-2em}
\begin{table}[H]
  \caption{Precision, Recall and Accuracy of the tools on DataRaceBench 1.2 including FP from True Race kernels excluding races due to \TT{pthread} mutex lock accesses}
\resizebox{0.8\columnwidth}{!}{
\begin{tabular}{ |l|c|c|c|c|c| }
\hline
Tools & Precision & Recall & Accuracy & F1 Score & Diagnostic odds ratio \\
\hline
\tool{Helgrind} & 0.24 & 0.95 & 0.31 & 0.38 & 2.77 \\
\hline
\tool{Valgrind DRD} & 0.24 & 0.95 & 0.31 & 0.38 & 2.80 \\
\hline
\tool{TSan-LLVM} & 0.26 & \textbf{0.97} & 0.27 & 0.41 & 4.48 \\
\hline
\tool{Archer} & \textbf{0.93} & 0.95 & \textbf{0.94} & \textbf{0.94} & \textbf{247.33} \\
\hline
\tool{SWORD} & 0.82 & 0.92 & 0.84 & 0.87 & 28.20 \\	\hline
\hline
\toolname & 0.70 & 0.96 & 0.79 & 0.81 & 41.14 \\
 \hline
\end{tabular}
}
\label{tbl:metrics-full-FP-in-TR}
\end{table}

\end{appendices}

\end{document}